# An Analysis of Optimal Link Bombs*

Sibel Adalı, Tina Liu, Malik Magdon-Ismail
Department of Computer Science, Rensselaer Polytechnic Institute
110 8th Street, Troy, NY 12180.
Email: {sibel, liut2, magdon}@cs.rpi.edu

November 21, 2018

**Abstract**

We analyze the phenomenon of collusion for the purpose of boosting the pagerank of a node in an interlinked environment. We investigate the optimal attack pattern for a group of nodes (attackers) attempting to improve the ranking of a specific node (the victim). We consider attacks where the attackers can only manipulate their own outgoing links. We show that the optimal attacks in this scenario are uncoordinated, i.e. the attackers link directly to the victim and no one else. nodes do not link to each other. We also discuss optimal attack patterns for a group that wants to hide itself by not pointing directly to the victim. In these disguised attacks, the attackers link to nodes $l$ hops away from the victim. We show that an optimal disguised attack exists and how it can be computed. The optimal disguised attack also allows us to find optimal link farm configurations. A link farm can be considered a special case of our approach: the target page of the link farm is the victim and the other nodes in the link farm are the attackers for the purpose of improving the rank of the victim. The target page can however control its own outgoing links for the purpose of improving its own rank, which can be modeled as an optimal disguised attack of 1-hop on itself. Our results are unique in the literature as we show optimality not only in the pagerank score, but also in the rank based on the pagerank score. We further validate our results with experiments on a variety of random graph models.

## 1 Introduction

Generally, a search for a particular topic on a particular search engine (such as *Google*) will output a ranked list of relevent web pages. The prominence of a page in this listing is an important indicator of how many people will visit the page. For a commercial web site, its prominence with respect to product searches has important financial consequences, as does the prominence of a competitor's website with respect to slander about products. Prominence in rankings is prestigious, can add credibility to a site or a concept and can be used to make political statements [16]. For example, a series of attempts, called Google bombs, to improve the ranking of certain sites for a specific keyword were used to give weight to a specific political point of view, e.g., making the web-biography of the U.S. President the top hit for the term "miserable failure"[1]. As a result of the importance

*A preliminary workshop version of this paper was presented at the First International Workshop on Adversarial Information Retrieval on the Web (AIRWeb 05) in conjunction with the 14th International World Wide Web Conference (WWW2005), Chiba, Japan, 10-14 May, 2005.

[1]The first Google bomb was with respect to the text "talentless hack". Since then several other attacks also succeeded in raising the ranks of web pages with respect to specific keyword(s), in some cases using as few as 25 links. It has been argued that several factors contributed to the success of these attacks: the number and prominence of the attacking pages; the (un)popularity of the keyword, the use of the same keyword in all links, the higher rankings of Blogs due the frequency of their updates, etc. Some of the keywords chosen in these attacks were very rare on the Web at the time of the attack: "French Military Victories". However, even attacks using keywords as popular as "Weapons of Mass Destruction" have been successful (BBC News, Sunday, 7 December, 2003).

attached to one's pagerank, especially one's Google pagerank, artificial methods for boosting one's pagerank are an active area for discussion. Pagerank is one of the many factors that is used in Google's ranking algorithm [18] and a significantly high pagerank can boost the prominence of a page considerably.

In addition to Google bombs that were oriented towards an external site, a web-retailer could also make use of link manipulation to improve the prominence of its own web-site with respect to a particular topic(s). Link farms are a common method for boosting pageranks [10] where a set of dummy pages are purposefully created to improve the pagerank of a specific page. However, in a link farm, the targeted page is controlled by the link farm as well. The link bombers or spammers are usually some (coordinated) set of web pages which add outgoing links to their web page. Some of these links will point to the attacked page, and contain the text they (the bombers) are trying to associate with the attacked page. The issue we address is how these bombers should organize their outgoing links in order to maximize the success of their link bomb in terms of pagerank score and rank.

There has been discussion on whether a link bomb can be considered an "undesirable" attack [20] that exploits a weakness in the pagerank-style algorithms [12, 18]. The pagerank algorithm assigns you a pagerank by considering the number and importance (according to PageRank) of web pages that point to you. Given that a search engine like Google currently ranks over 10 billion pages, one would expect that a very small number of web pages should not be able to change the ranking of a page dramatically, contrary to what has been observed. Thus, one motivation for studying the optimal attack is to determine specific abnormal but effective attack patterns that could be identified as artificial link bombs.

We present results on the optimal link bomb. Specifically, the *attackers* are a set of web pages whose outgoing links can be manipulated, and the *victim* is the target web page to be bombed. THe victim's outgoing links cannot be manipulated. Our main result is to establish the following theorems as a starting point for a discussion of accountability on linked structures such as the WWW.

**Theorem 1.** The attack which maximizes the **pagerank score** of the victim is the direct individual attack.

**Theorem 2.** The attack which maximizes the **rank** of the victim is the direct individual attack.

Rank is the order statistic defined by the pagerank, and the *direct individual attack* is the attack in which every attacker points *only* to the victim and to no other page. In particular, in the optimal attack, none of the attackers point to each other. Thus, the optimal attack masquerades as a set of uncoordinated "random" nodes, all pointing to the same page. Note that both the stated theorems are non-trivial. An attack that maximizes pagerank score of the victim is not necessarily one that maximizes the pagerank rank, if the attack also raises some other node's pagerank score above the victim's score. To our knowledge, our result is the first result in terms of rank.

We also discuss optimal "disguised" attack patterns, in which none of the attackers wish to directly point to the victim – all paths from the attackers to the victim must be of at least some minimum length $\ell$ from the victim. In this case the optimal attack is still a direct individual attack, however now the attackers point to some other *intermediate node* (not the victim).

**Theorem 3.** *There is an optimal disguised attack in which every attacker's only link is to the same node which is distance $\ell - 1$ from the victim.*

In our work, we assume that the attackers can only control their outgoing links. They may not control the outgoing links of any other nodes (so the target page is outside the attacking set). Here, we give the extensions and complete proofs of optimality of the original results on optimal direct and disguised attacks, which were initially discussed in preliminary form in [1]. In disguised attacks, *no* attacker can point to the attacked node. For malicious link bombs, it is reasonable to assume that the target is outside the attacker set. When the attackers are trying to boost one of their own sites, however, the attackers can control the outgoing links of the target page. This is the case in link farms. The optimal configuration in a link farm therefore follows directly from our results: in the link farm all the attackers except the target use the direct attack to improve the target's rank. The target (who is now also an attacker) uses the optimal disguised attack of length $\ell = 1$, after the other attackers have made their direct attacks to boost its own rank.

While the optimal attack is always the direct individual attack, the amount by which the direct individual attack surpasses other (more coordinated) attack patterns may depend on the nature of the graph. We give experimental results that quantify this phenomenon for a variety of different attack patterns. On certain random graph models of the Web, some coordinated attack patterns are almost as good as the direct individual attack, and can hence be used in place of the direct individual attack as a means of disguising the attack. While the effect of graph structure on the pagerank has been investigated in the literature [17, 12], to our knowledge, these are the first results regarding the effect of the graph structure on the effectiveness of link bombs.

Our results raise interesting questions such as how to detect and respond to link bomb attacks (in general this problem is NP-hard, see for example [22]). Since the attackers will have no visible associations amongst themselves, it is hard to detect and prove that they are participating in an attack. If the optimal attack were a tree structure, there would be a small set of nodes with high prominence that one might argue are "responsible" for the attack. The other nodes pointing to these nodes could also be held accountable aiding and abbeting the actions of the responsible nodes. Such accountability is not possible in an individual attack.

We proceed by first discussing the related work and giving some preliminary definitions, followed by a preview of our result for an isolated graph, in which the only nodes are the attackers and the victim. We then discuss general graphs, followed by some experimental results on a variety of random graph models. We conclude with a discussion of the implications of our results. (We defer some technical proofs to an appendix)

## 1.1 Related Work

Link spam has received significant attention recently, and most of the work goes along the lines of quantifying the impact of different collusion strategies on pagerank [13, 2, 7, 10]. Bianchini [4] analyzes the impact of different community structures in the optimal energy, i.e. total pagerank value for a set of pages. Another line of research concentrates on the problem of modifying pagerank to make it resistant to such collusion strategies [14, 23, 6]. In particular, [14] concentrates on using a set of handpicked trusted sites to bias the pagerank computation and develops methods for selecting seeds to be evaluated in this algorithm. Similarly, Zhang et. al. [23] develop a method for stalling the random jump probabilities to reduce the impact of colluding web pages. Caverlee et. al. [6] introduce the notion of domain or host level influence throttling to combat link spam. Drost and Scheffer [9] introduce machine learning algorithms to recognize spam pages, including those with link spam; their work considers both number of incoming and outgoing links as well as features related to the content.

We highlight two of the works which are the most closely related to ours. The work by Gyöngyi and Garcia-Molina [13] was developed independently of ours and has a similar flavor. In particular, they consider the case of optimal link spam structure under the assumption of constant leakage, which is a significant limitation. Additionally, they compute the magnitude of the attacks for various attack patterns. The limitation of the constant leakage was addressed by Du, Shi and Zhao in [10]. In particular, they consider the possibility that the attackers can have control of other pages in addition the link spam farm. Du, Shi and Zhao also consider disguised attacks, when the attacking nodes must point to non-target nodes (in addition to the possibility of pointing to target nodes). One difference between this existing work and ours, is that it is focussed on the pagerank bombing. We do provide results for pagerank bombing, but our main result is to show that the same optimality of the direct attack holds for the rank, which is more difficult to analyze. However, we do not quantify algebraically the improvement in pagerank scores for different link farm configurations, which is shown in Du, Shi and Zhao and Gyöngyi and Garcia-Molina.

Our results for optimal disguised attacks can be converted to algorithms, however these algorithms require global knowledge of the graph to implement, which may be non-realistic for bounded complexity attackers. Du, Shi and Zhao [10] also consider disguised attacks, but allow the attackers to point to the target but in a non-obvious way; an optimal strategy chooses nodes to point to so as to minimize the leakage in pagerank forced by the disguise. In general, computing any optimal disguised attack should involve the knowledge of the entire graph. An interesting open question is whether there are near-optimal disguised attacks which can be locally computed, only knowing some bounded in and out-neighborhood of the target and similarly some bounded in-neighborhood of the attackers.

## 2 Preliminaries

A search query on a set of keywords results in an ordered list of web pages $\mathcal{W} = \{\omega_i\}$. Each web page $\omega \in \mathcal{W}$ contains some or all of the keywords either in its text or in the text of a link that points from some other web page to $\omega$. A scoring function is used to order the pages in $\mathcal{W}$. The most prominent page (page with the highest score) is given rank 1, etc.

Google [5] considers many factors in its scoring function, including: keyword frequency; relative locations of the keywords; the position and style of the keywords. An important factor in the scoring function is the *pagerank* which depends on how the web page is embedded in the entire graph of web pages. An early paper on the Google system [5] suggests that no one factor dominates the scoring function, however, the pagerank plays an important role. In this paper, we will concentrate only on the pagerank factor and discuss how it can be manipulated.

The *web graph* is a directed graph $G = (V, E)$ that models the World Wide Web. The vertex set $V$ represents the pages and documents, and the edge set $E$ represents the links between the pages and documents[2]. The edges are directed: if $(v_1, v_2) \in E$, then $v_1$ contains a link to $v_2$. In a web graph, the in-degree $indeg(v)$ of page $v$ is the number of links that point to $v$ and the out-degree $outdeg(v)$ is the number of links originating from $v$ that point to other pages. A (directed) *path* of length $\ell$ is a sequence of vertices $v_0, v_1, \ldots, v_\ell$ with $(v_{i-1}, v_i) \in E$ for $i = 1, \ldots, \ell$. $v_\ell$ is the terminal node in the path, and $v_1, \ldots, v_{\ell-1}$ are intermediate nodes. We allow parallel edges between two vertices, but no self-loops.

The pagerank $p_i$ models the probability that node $i$ will be visited either by randomly navigating down links in the web graph or by randomly jumping to page $i$. Let $\alpha$ be the probability to navigate, and $1-\alpha$ the probability to jump. Then the pageranks $\{p_j\}$ of the nodes in a graph simultaneously satisfy the set of linear equations[3]

$$p_i = \alpha \sum_{(v_j, v_i) \in E} \frac{p_j}{outdeg(v_j)} + \frac{1-\alpha}{N}. \tag{1}$$

($0 \leq \alpha \leq 1$ and $N = |V|$.) The first term represents the probability to reach $i$ by random navigation. An edge may appear multiple times if there are parallel links. The second term represents the probability to reach $i$ by randomly jumping. Typically, $\alpha \in [0.85, 0.95]$. The pagerank $p_i$ is larger if $v_i$ has a large in-degree, and its incoming links are from high pagerank nodes with small out-degree. The PageRank algorithm [18] is an iterative approach to solving these equations. The pageranks are all initialized to $p_i^0 = \frac{1}{N}$. The PageRank iteration is given by

$$p_i^{t+1} = \alpha \sum_{(v_j,v_i)\in E} \frac{p_j^t}{outdeg(v_j)} + \frac{1-\alpha}{N}. \tag{2}$$

[2]Note that the definition of an edge is traditionally given by hyperlinks in a web page. However, it is also possible to count URLs in the body of a web page as links. The definition of what constitutes a link is usually application dependent.

[3] An alternative and common formulation of the pageranks in the literature is as the stationary distribution of a suitably defined finite irreducible Markov chain with transition matrix $P = (1-\alpha)M + \alpha U$, where $U$ is a matrix of 1's. Many of our results could be obtained by analyzing how the stationary distribution changes under perturbations of $P$. Our approach is more graph theoretic, treating the problem as a flow.

$p_i^t$ converges to the (unique) solution of (1). We assume that every page can manipulate its outgoing links, but it cannot change its incoming links.

A *link bomb*, or *attack* occurs when a group of *attackers* $A = \{v_1, \ldots, v_K\}$ alter their outgoing links so as to boost the pagerank of a *victim* $v_0 \notin A$. Before the attack, if the edge set is $E$, then after the attack the edge set will be $\bar{E}$ where the only edges added or removed from $E$ are of the form $(v_i, u)$ where $1 \leq i \leq K$ and $u \in V$, i.e., the attackers may remove and/or add outgoing links only. After the attack, the new web graph is $\bar{G} = (V, \bar{E})$. Let $p_i$ denote the pageranks in the original graph $G$ (before the attack), and $\bar{p}_i$ the pageranks in $\bar{G}$ (after the attack). The *magnitude* of the attack $\Delta p_0 = \bar{p}_0 - p_0$ is the amount by which the pagerank of the victim increased, and is a measure of the success of the attack. In our analysis, we only consider the magnitude of the attack, and assume that all other factors entering into the scoring function are unchanged.

# 3 The Optimal Link Bomb

In this section, we investigate how to maximize the magnitude of the attack. In particular, we show that the effectiveness of the attack *does not* increase if the attackers try to coordinate the attack in some way, by introducing links among themselves in order to increase their ranks. (Recall that, incoming links from higher ranked pages are more beneficial to your rank.) First, we consider a simplified case, in which the attackers and the victim are isolated from the rest of the graph. We then consider the general case.

## 3.1 Isolated Graphs

We first restrict our attention to a graph whose vertex set is composed only of the attackers and the victim, $V = A \cup v_0$ (i.e., $N = |V| = K + 1$). Assume (for simplicity) that $v_0$ does not point to any member of $A$. We first consider some examples of attacks, before giving the general result. In all cases, all the attackers in $A$ point to the victim $v_0$, and what differentiates the attacks is how the attackers are themselves organized.

*Direct Individual:* The only links are to $v_0$.

*Tree:* The attackers form a tree. For any graph with a topological order, one can compute the pageranks efficiently (in linear time). We will specialize to a *star attack* in which $v_2 \ldots, v_K$ point to $v_1$ and all attackers point to $v_0$.

*Cycle:* The attackers form a cycle.

*Complete:* The attackers a complete graph.

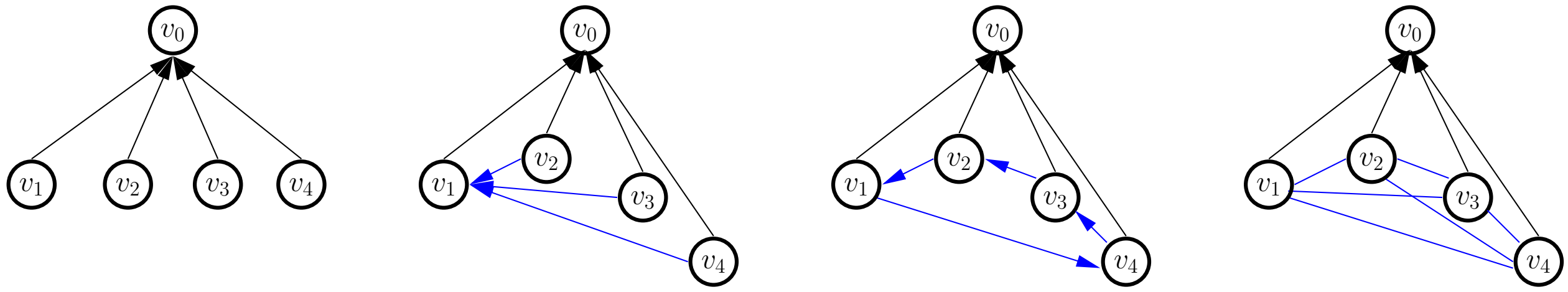

Direct Individual | Tree | Cycle | Complete

By solving the linear system (1) for the graph resulting from each of these attacks, we obtain

**Lemma 4.** *For the isolated graph,*

$$\begin{aligned}\bar{p}_0(\textit{individual}) &= p_0(1+\alpha K),\\ \bar{p}_0(\textit{star}) &= p_0\left(1+\tfrac{\alpha}{2}(K(1+\alpha)+1-\alpha)\right),\\ \bar{p}_0(\textit{cycle}) &= p_0\left(1+\tfrac{\alpha K}{2-\alpha}\right),\\ \bar{p}_0(\textit{complete}) &= p_0\left(1+\tfrac{\alpha K}{K(1-\alpha)+\alpha}\right),\end{aligned}$$

*where* $p_0 = (1-\alpha)/(K+1)$ *is the initial pagerank of* $v_0$.

Since $0 \le \alpha \le 1$, after some algebra, we obtain

**Theorem 5.** *For the isolated graph,*

$$\bar{p}_0(\textit{individual}) \ge \bar{p}_0(\textit{star}) \ge \bar{p}_0(\textit{cycle}) \ge \bar{p}_0(\textit{complete}).$$

We will show that the direct individual attack is optimal for the isolated graph. Note that when a node has zero outdegree, it "stunts" the flow of pagerank. This means that the sum of the pageranks need not be 1, i.e. $\{p_i\}$ need not be a probability distribution. Summing (1) over $i$, we get

$$\sum_i p_i = \alpha \sum_i \sum_{(v_i,v_j)\in E} \frac{p_j}{\textit{outdeg}(v_j)} + 1 - \alpha.$$

If $\textit{outdeg}(v_i) > 0$, $v_i$ contributes $\frac{p_i}{\textit{outdeg}(v_i)}$ exactly $\textit{outdeg}(v_i)$ times to the summation for a total contribution of $p_i$. If $\textit{outdeg}(v_i) = 0$, then $v_i$ does not contribute to the summation, so we obtain

$$\begin{aligned}\sum_i p_i &= \alpha \sum_{\textit{outdeg}(v_i)>0} p_i + 1 - \alpha,\\ &= \alpha \sum_i p_i + 1 - \alpha - \alpha \sum_{\textit{outdeg}(v_i)=0} p_i.\end{aligned}$$

After rearranging terms and solving for $\sum_i p_i$, we obtain the following useful lemma.

**Lemma 6.** $\displaystyle\sum_i p_i = 1 - \frac{\alpha}{1-\alpha} \sum_{\textit{outdeg}(v_j)=0} p_j.$

This lemma is useful for proving the next theorem; though it is a special case of the general result in the next section, it is illustrative and the proof gives an intuition for the general case.

**Theorem 7.** *For an isolated graph, the individual attack uniquely maximizes* $p_0$.

*Proof.* Since $\sum_{i=1}^{K} p_i = \sum_{i=0}^{K} p_i - p_0$, Lemma 6, gives

$$\begin{aligned}\sum_{i=1}^{K} p_i &= 1 - \frac{\alpha}{1-\alpha} \sum_{outdeg(v_i)=0} p_i - p_0, \\ &\leq 1 - \frac{p_0}{1-\alpha},\end{aligned}$$

with equality *iff* $v_0$ is the only vertex with degree 0. For an arbitrary attack,

$$\begin{aligned}p_0 &= \alpha \sum_{(v_i,v_0)\in E} \frac{p_i}{outdeg(v_i)} + \frac{1-\alpha}{K+1}, \\ &\overset{(a)}{\leq} \alpha \sum_{(v_i,v_0)\in E} p_i + \frac{1-\alpha}{K+1}, \\ &\overset{(b)}{\leq} \alpha \left(1 - \frac{p_0}{1-\alpha}\right) + \frac{1-\alpha}{K+1}.\end{aligned}$$

Solving for $p_0$, we obtain that

$$p_0 \leq \left(\frac{1-\alpha}{K+1}\right)(1+\alpha K).$$

This bound is attained by the individual attack. Uniqueness follows because equality in (a) occurs *iff* $outdeg(v_i) = 1$ whenever $(v_i, v_0) \in E$, and equality in (b) occurs *iff* every edge $(v_i, v_0)$ is in $E$. ■

## 3.2 Arbitrary Graphs

When $v_0, \ldots, v_K$ are embedded in a larger graph $G$, the direct individual attack is still optimal. Intuitively, one can view the PageRank iteration (2) as sending a flow of pagerank down the directed edges. The maximum flow from $v_i$ to $v_0$ occurs when $v_i$ points directly to $v_0$, and to no other node – any other links divert the flow and leads to a lower magnitude attack. The following results will make this intuition more formal. We will generally refer to nodes which are neither the attackers nor the victim by $w_j$, and $u_j$ will be used to refer to any node. The 1-neighborhood $N_1(v)$ of a node $v$ is the set of nodes to which $v$ points. $N_k(v)$ $(k > 1)$ is the set of $k$-neighborhood nodes: $u \in N_k(v)$ *iff* for some $w \in N_{k-1}(v)$, $(w, u) \in E$. Note that $v$ could be in its own $k$-neighborhood for $k > 1$, and $N_0(v) = \{v\}$. In this section, many of the proofs are involved, and so we will sketch the intuition and defer the technical proofs to the appendix.

Consider attacker $v_i$, and, without loss of generality, assume it initially has no outgoing links. Suppose now that it adds $\delta$ outgoing edges. This results in $\frac{\alpha}{\delta}$ of its rank "flowing" along each of its edges to its neighbors (note there may be parallel links). Thus, the rank increase for a 1-neighbor $u_j$ is given by

$$\Delta_j^1 = \alpha \sum_{(v_i,u_j)\in E} \frac{p_i}{outdeg(v_i)},$$

where the superscript 1 indicates that $u_j$ is a 1-neighbor, and $j$ is an index that enumerates the 1-neighbors. The sum is over all parallel edges that $v_i$ may have to $u_j$. This increase in rank in turn propagates to 2-neighbors, resulting in an increase in the rank of a 2-neighbor $u_k$ by an amount

$$\Delta_k^2 = \alpha \sum_{\substack{(u_j,u_k)\in E \\ s.t\ u_j\in N_1(v_i)}} \frac{\Delta_j^1}{outdeg(u_j)}.$$

The sum is over all 1-neighbors pointing to $u_k$ (including parallel edges). If the newly added edges create a path from $v_i$ to $v_0$, then some amount of $v_i$'s pagerank will propagate to $v_0$. We define $\Delta_j^l$ to be the change in the page rank of $u_j$ from flow down all paths of length $l$ from $v_i$ to $u_j$,

$$\Delta_j^l \;=\; \alpha \sum_{\substack{(u_k,u_j)\in E\\ s.t\ u_k\in N_{l-1}(v_i)}} \frac{\Delta_k^{l-1}}{outdeg(u_k)}$$

Let $\delta(l)$ be the total increase in page rank through paths of length $l$, $\delta(l) = \sum_j \Delta_j^l$. Since the pagerank increase attenuates by a factor $\alpha$ with each edge, we have the following lemma.

**Lemma 8.** *For $\ell \geq 1$,*

$$\delta(l) \leq \alpha^l p_i,$$

*with equality* iff $\delta(l-1) = \alpha^{l-1}p_i$ *and for every* $u_k \in N_{l-1}(v_i)$, $outdeg(u_k) > 0$.

*Proof.* See Section 4.1.1. ∎

Let $\mathcal{S}$ be a set of nodes. A path $q$ passes through $\mathcal{S}$ if some node of $\mathcal{S}$ is an intermediate node of $q$. A set of paths $P$ pass through $\mathcal{S}$ if every path in $P$ passes through $\mathcal{S}$. Let $P_t$ be a collection of paths that passes through $\mathcal{S}$, with every path in $P_t$ having the same terminal node $t \neq v_i$ ($t$ is not an intermediate node of any path in $P_t$). We call $t$ a *progeny* of $\mathcal{S}$ with respect to the paths $P_t$. Since every path passes through $\mathcal{S}$, some prefix of every path in $P_t$ has a terminal node in $\mathcal{S}$. For each path $q \in P_t$, let $q_{\mathcal{S}}$ be a (any) prefix with terminal node in $\mathcal{S}$, and let $P_t(\mathcal{S})$ denote the collection of such distinct prefixes $\{q_{\mathcal{S}}\}$.

The *influence* $I(\mathcal{S}|P_t(\mathcal{S}))$ of $v_i$ on $\mathcal{S}$ is the total flow of pagerank (summed over all nodes in $\mathcal{S}$) from $v_i$ to $\mathcal{S}$ along the paths in $P_t(\mathcal{S})$ (which are (distinct) prefixes in $P_t$). The influence $I(t|P_t)$ of $v_i$ on $t$ is the total flow of pagerank that flows to $t$ along the paths in $P_t$ (which pass through $\mathcal{S}$). Every path in $P_t$ has at least one additional edge compared with its corresponding prefix that terminates in $\mathcal{S}$, so the influence that propagates to $t$ along $P_t$ can be at most the influence that propagates to $\mathcal{S}$ along the paths in $P_t(\mathcal{S})$, attenuated by a factor $\alpha$. We have the following lemma.

**Lemma 9.** *Let $P_t$ be a collection of paths from $v_i$ to $t$ which passes through a set of nodes $\mathcal{S}$ ($t$ appears only as a terminal node in $P_t$). Let $P_t(\mathcal{S})$ be a (any) collection of distinct prefixes terminating in $\mathcal{S}$. Then*

$$I(t|P_t) \leq \alpha I(\mathcal{S}|P_t(\mathcal{S})),$$

*independent of which prefixes are used in the construction of $P_t(\mathcal{S})$.*

*Proof.* See Section 4.1.2. ∎

We now consider $v_i$'s attack on $v_0$. Let $P$ denote the collection of all (distinct) paths from $v_i$ to $v_0$ in which $v_0$ appears only as the terminal node, i.e., $v_0$ is not an intermediate node of any path in $P$. Note that if there are cycles in the graph, then $P$ may contain an infinite number of paths. Let the flow of pagerank from $v_i$ to $v_0$ down the paths in $P$ be denoted $\Delta$. There may be cycles containing $v_0$, in which case, the pagerank increase $\Delta$ will continue to flow around these cycles, back to $v_0$ increasing the pagerank further, i.e., $\Delta$ will be amplified by the cycles. Let $\Delta p_0^i$ be $v_i$'s contribution to the magnitude of the attack,

$$\Delta p_0^i(\Delta) = \Delta + \text{amp}(\Delta),$$

where amp($\Delta$) is the amplification due to the cycles that contain $v_0$. The larger $\Delta$, the larger will be the amplification of $\Delta$,

**Lemma 10.** $\Delta p_0^i(\Delta)$ *is a monotonically increasing function of* $\Delta$.

*Proof.* See Section 4.1.3. ■

Lemmas 8, 9 and 10 are the main tools we will need to prove our main result, namely that the individual attack is optimal with respect to pagerank. By Lemma 10, since $\Delta p_0^i$ is monotonically increasing in $\Delta$, $\Delta p_0^i$ will be maximized when $\Delta$ is maximized. $\Delta$ is given by the sum of the flows of pagerank from $v_i$ to $v_0$ along the paths in $P$, therefore we only need to consider this flow.

Let $\ell$ be the length of the shortest path in $P$ (there may be many such shortest paths). Consider the set $L$ of all distinct paths of length $\ell$ originating at $v_i$. Some of these paths have terminal node $v_0$. We now restrict our attention to the set $L'$ containing those paths in $L$ which do not have terminal node $v_0$. Note that none of the paths in $L'$ can have $v_0$ as an intermediate node since the shortest path from $v_i$ to $v_0$ has length $\ell$. Let $\mathcal{S}$ denote the set of terminal nodes in $L'$. Partition $P$ into two disjoint sets, $P_\ell$ and $P_{>\ell}$, where $P_\ell$ contains the paths in $P$ with length $\ell$ and $P_{>\ell}$ the paths with length $> \ell$. Every path in $P_{>\ell}$ must pass through at least one of the nodes in $\mathcal{S}$, therefore $P_{>\ell}$ passes through $\mathcal{S}$. Every path in $P_{>\ell}$ has terminal node $v_0$, and $v_0$ does not appear as an intermediate node in any of these paths. Thus, $v_0$ is a progeny of $\mathcal{S}$ with respect to $P_{>\ell}$. Every path in $P_{>\ell}$ has a prefix of length $\ell$ with terminal node in $\mathcal{S}$. Collect these distinct prefixes into the set $P_{>\ell}(\mathcal{S})$.

Let $\Delta_\ell$ be the contribution to $\Delta$ due to flow along the paths in $P_\ell$, and $\Delta_{>\ell}$ the contribution due to flow along the paths in $P_{>\ell}$. Then,

$$
\begin{aligned}
\Delta &= \Delta_\ell + \Delta_{>\ell}, \\
&\overset{(a)}{=} \Delta_{v_0}^\ell + I(v_0|P_{>\ell}), \\
&\overset{(b)}{\leq} \Delta_{v_0}^\ell + \alpha I(\mathcal{S}|P_{>\ell}(\mathcal{S})), \\
&\overset{(c)}{\leq} \Delta_{v_0}^\ell + I(\mathcal{S}|P_{>\ell}(\mathcal{S})), \\
&\overset{(d)}{\leq} \Delta_{v_0}^\ell + \sum_{s\in\mathcal{S}} \Delta_s^\ell \\
&\overset{(e)}{\leq} \delta(\ell) \\
&\overset{(f)}{\leq} \alpha^\ell p_i
\end{aligned}
$$

(a) follows from the definitions of $\Delta_{v_0}^\ell$ and influence; (b) follows from Lemma 9 and (c) because $\alpha \leq 1$. (d) follows because the paths in $P_{>\ell}(\mathcal{S})$ are all of length $\ell$, so $P_{>\ell}(\mathcal{S})$ is a subset of all the paths of length $\ell$ that terminate in $\mathcal{S}$; (e) follows from the definition of $\delta(\ell)$, since $\mathcal{S} \cup v_0 \subseteq N_\ell(v_i)$; finally, (f) is an application of Lemma 8. Equality occurs *iff* $\mathcal{S}$ is empty, and all paths from $v_i$ are of length $\ell$, ending at $v_0$. Certainly, the optimal value of $\ell$ is 1, and so we have the following theorem[4].

**Theorem 11.** *$\Delta p_0^i$ is maximized if and only if the* only *edge from $v_i$ is to $v_0$. This is independent of all the other edges in the graph, in particular independent of the edges from the other $v_j$.*

Theorem 11 directly implies the following result,

**Corollary 12.** *The direct individual attack is optimal for maximizing the pagerank $p_0$.*

Though the direct individual attack maximizes the pagerank of $v_0$, it is not obvious that this also maximizes the rank of $v_0$, which depends on the relative pageranks. Is it possible that some other attack, though it will increase $p_0$ less, might increase it more relative to some other node and hence improve $v_0$'s rank more? The answer is no, i.e. the direct direct individual attack also maximizes the rank (as opposed to the pagerank) of the victim.

[4] An alternative proof of this theorem using the Markov chain approach can be given using a generalization of the result in [8], where it is shown that adding the edge $(i, j)$ can only increase the pagerank of $j$.

Suppose that some other attack $X$ maximizes the rank of $v_0$. This means that for some node $u$, $\bar{p}^I_{v_0} \le \bar{p}^I_u$ and $\bar{p}^X_{v_0} > \bar{p}^X_u$ ($I$ denotes the direct individual attack). We show that such a situation can never occur, leading to the following result.

**Theorem 13.** *The direct individual attack maximizes the rank of $v_0$.*

*Proof.* See Section 4.1.4. ∎

### 3.3 The Optimal Disguised Attack

We now consider the situation in which the attackers wish to maximize the magnitude of their attack on $v_0$, but they wish to disguise the attack by not pointing directly to the victim. In such an attack, the anchor text will not be associated to the victim, hence we assume that the victim already has a high prominence with respect to the anchor text. The specific disguise constraint we consider is that for every attacker, the shortest path to the victim should have length at least $\ell \ge 1$.

Consider attacker $v_i$. In any attack, some amount of pagerank flows from $v_i$ to $v_0$. In any directed graph, we define $f(u;v)$, the *forward value* of vertex $u$ with respect to vertex $v$, to be the fraction of $u$'s pagerank that flows to $v$ along paths with $v$ as terminal node but not as intermediate node. Thus, for example, $f(v;v) = 1$. Since the fraction of $u$'s rank that makes it to $v$ can be obtained by multiplying the fraction flowing to each neighbor with the fraction flowing from that neighbor to $v$, we obtain the *forward equation* for the forward values $f(u;v)$:

$$\begin{aligned} f(v;v) &= 1, \\ f(u;v) &= \frac{\alpha}{outdeg(u)} \sum_{(u,w)\in E} f(w;v). \end{aligned} \tag{3}$$

The forward equation (3) is similar to the pagerank equation (1) and can be solved by a similar iterative algorithm as in (2).

For every vertex $u$ (not an attacker), we consider the edge set $E_u = E \cup (v_i, u)$, which defines a new directed graph in which the edge set is augmented by a single link from the attacker to $u$. For this graph, we can compute the forward value $f_u(w;v_0)$ of any vertex $w$ with respect to $v_0$. We define the value $V_i(u)$ of vertex $u$ to attacker $v_i$ by

$$V_i(u) = f_u(v_i; v_0).$$

By Lemma 10 the optimal attack is the one that maximizes the flow of pagerank to $v_0$, which means that $v_i$ should point to the node $u$ satisfying the "disguise constraints" that maximizes $V_i(u)$. There may be many optimal attacks, but we will now show that there exists an optimal attack for $v_i$ which consists of adding a *single* link to the vertex $u$ that maximizes $V_i(u)$, which is at distance $\ell - 1$ from $v_0$. Let $d(u,v)$ be the length of the shortest path from $u$ to $v$; if no path exists from $u$ to $v$, set $d(u,v) = \infty$. Let $U_l(v_0)$ be the collection of nodes which have a path of length $l$ to $v_0$ and no shorter path to $v_0$. Thus,

$$U_l(v_0) = \{u : d(u, v_0) = l\}.$$

Suppose that the disguise constraint (which we apply to all the attackers) is that the shortest path from an attacker to $v_0$ must have length at least $\ell$. Let $U_{\ell-1} = U_{\ell-1}(v_0)$ be the nodes with a path of length $\ell - 1$ to $v_0$. First we show that the maximum value of $V_i(u)$ is attained for some node in $U_{\ell-1}$.

**Lemma 14.** $\max\limits_{u:d(u,v_0)\ge \ell-1} V_i(u) = \max\limits_{u\in U_{\ell-1}} V_i(u)$.

*Proof.* See Section 4.2.1. ■

Lemma 14 implies that we only need consider nodes that are distance $\ell - 1$ to $v_0$ in determining which intermediate node to attack. Note that for each $u \in U_{\ell-1}$, in order to compute $V_i(u)$, we need to compute $f_u(v_i, v_0)$, which may require the computation of $f_u(v, v_0)$ for all $v \in V$. By following arguments similar to those that led to Theorem 11, we find that the optimal attack for $v_i$ is to point only to the vertex $u$ that maximizes $V_i(u)$.

**Theorem 15.** *The optimal disguised attack for a single attacker $v_i$ is a* single *link to the vertex $u$, at distance $\ell - 1$ from $v_0$, which maximizes $V_i(u)$.*

*Proof.* See Section 4.2.2. ■

Note that the vertex that maximizes $V_i(u)$ may not be unique, however by Lemma 14, we know that at least one such vertex exists in $U_{\ell-1}$.

Unfortunately, the maximizing node $V_i(u)$ need not be the same for different attackers – the disguise constraint introduces dependencies between attackers, i.e., the optimal attack for a particular attacker may depend on what the other attackers do. In particular, it is no longer the case that each attacker using its optimal disguised individual attack will maximize the magnitude of the disguised attack if the group of attackers act jointly. The following example with two attackers and $\ell = 2$ illustrates the issue.

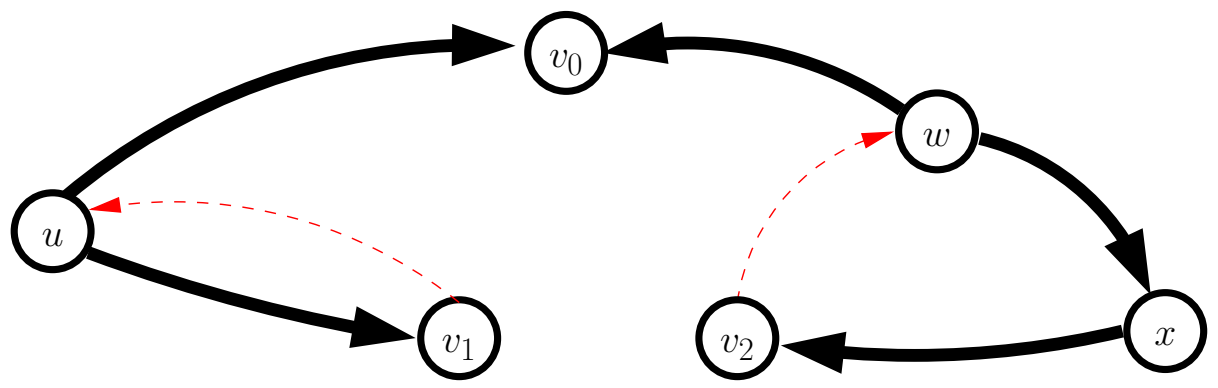

(a) Optimal individual attacks.

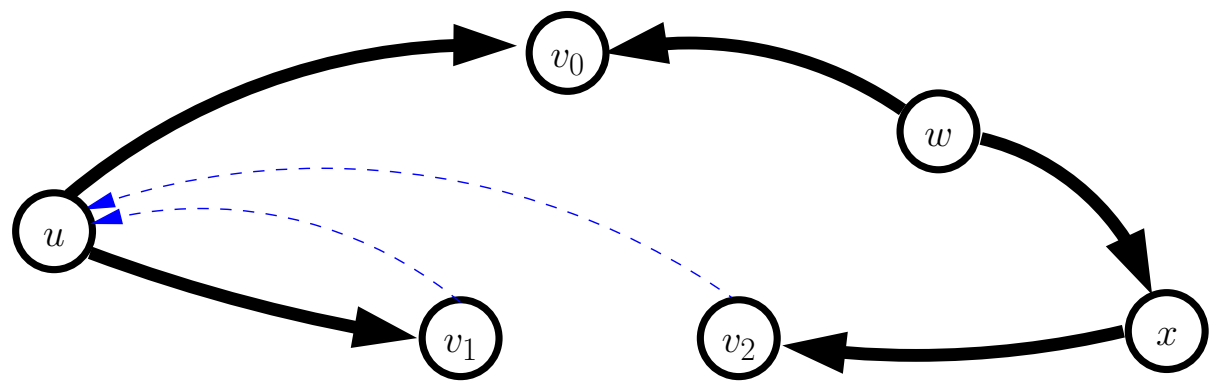

(b) Optimal joint attack.

The optimal attack for $v_1$ is to point to $u$, and for $v_2$ it is to point to $w$ (red dotted arrows in (a)). However, if both attackers attack, then they should both point to $u$. Theorem 15 applies to attacker $v_i$, independently of what the other attackers do. In particular, we conclude that in the optimal joint attack, every attacker has a single link to a node in $U_{\ell-1}$. In fact, there is an optimal attack in which every attacker links to the *same* node in $U_{\ell-1}$,

**Theorem 16.** *There is an optimal joint attack in which every attacker points to the same node in $U_{\ell-1}$.*

*Proof.* See Section 4.2.3. ■

Theorem 16 ensures that an efficient algorithm to compute an optimal joint attack is to select the best attack among all the attacks in which the attackers all link to a single node in $U_{\ell-1}$ (there are at most $O(|V|)$ such attacks).

# 4 Proofs

For our proofs, we will need some standardized notation for discussing sets of paths, and flow of pagerank along these paths. A collection of paths $P(w_1 w_2; x_1 x_2 \cdots x_k)$ contains all paths from $w_1$ to $w_2$ which do not contain the nodes $x_1, \ldots, x_k$ as intermediate nodes. The fraction of $w_1$'s pagerank that flows to $w_2$ along the paths in $P(w_1 w_2; x_1 x_2 \cdots x_k)$ will be denoted $\rho(w_1 w_2; x_1 x_2 \cdots x_k)$. Since only positive flow flows along paths, we have the following useful lemma,

**Lemma 17.** *If $S_1 \subseteq S_2$ are two sets of nodes, then $\rho(w_1 w_2; S_1) \geq \rho(w_1 w_2; S_2)$.*

Consider cycles originating at a node $w$ and not containing $w$ as an intermediate node. Suppose that a fraction $\gamma$ of $w$'s page rank flows along these cycles back to $w$. Since this fraction can also flow back to $w$ along the same cycles (attenuated by an additional $\gamma$ factor), by summing the resulting geometric series, we obtain the following useful lemma,

**Lemma 18.** *Consider a node $w$ and a set of nodes $S$ with $w \in S$. Let $\gamma = \rho(ww; S)$. Then, the fraction of $w$'s pagerank that flows back of $w$ through repeated use of the cycles in $P(ww; S)$ is* $\dfrac{1}{1-\gamma}$.

## 4.1 Arbitrary Graphs

### 4.1.1 Proof of Lemma 8

We prove the lemma by induction on $l$. When $l = 1$, if $outdeg(v_i) = 0$ then $\delta(1) = 0 \le \alpha p_i$. If $outdeg(v_i) > 0$, then

$$\begin{aligned}
\delta(1) &= \sum_{u_j} \alpha \sum_{(v_i,u_j)\in E} \frac{p_i}{outdeg(v_i)}, \\
&= \frac{\alpha p_i}{outdeg(v_i)} \sum_{u_j} \sum_{(v_i,u_j)\in E} 1, \\
&\overset{(a)}{=} \alpha p_i.
\end{aligned}$$

(a) follows because $\sum_{u_j} \sum_{(v_i,u_j)\in E} 1 = outdeg(v_i)$. Thus, $\delta(1) \le \alpha p_i$. Suppose that $\delta(L) \le \alpha^L p_i$, and consider $l = L + 1$.

$$\begin{aligned}
\delta(L+1) &= \sum_{u_j \in N_{L+1}(v_i)} \Delta_j^{L+1}, \\
&= \sum_{u_j} \alpha \sum_{\substack{(u_k,u_j)\in E \\ s.t.\ u_k \in N_L(v_i)}} \frac{\Delta_k^L}{outdeg(u_k)}, \\
&= \alpha \sum_{\substack{u_k \in N_L(v_i) \\ s.t.\ outdeg(u_k)>0}} \frac{\Delta_k^L}{outdeg(u_k)} \sum_{u_j} \sum_{(u_k,u_j)\in E} 1, \\
&\overset{(a)}{=} \alpha \sum_{\substack{u_k \in N_L(v_i) \\ s.t.\ outdeg(u_k)>0}} \Delta_k^L, \\
&\overset{(b)}{\le} \alpha \sum_{u_k \in N_L(v_i)} \Delta_k^L, \\
&\overset{(c)}{=} \alpha \delta(L), \\
&\overset{(d)}{\le} \alpha^{L+1} p_i.
\end{aligned}$$

(a) follows because $\sum_{u_j} \sum_{(u_k,u_j)\in E} 1 = outdeg(u_k)$. Equality in (b) occurs only if all nodes $u_k \in N_L(v_i)$ have $outdeg(u_k) > 0$. (c) follows from the definition of $\delta(L)$, and (d) from the induction hypothesis. Equality in (d) occurs only if $\delta(L) = \alpha^L p_i$. Thus the claim holds for all $l > 0$, which together with the conditions for equality concludes the proof of the theorem. ■

Since every link in a path attenuates the pagerank flow by at least $\alpha$, we have the following lemma, which will be useful in the proof of Lemma 9.

**Lemma 19.** *For any two nodes $u$ and $v$, not necessarily distinct, and any set of nodes $S$ containing $v$, $\rho(uv;S) \le \alpha^\ell$, where $\ell$ is the length of the shortest path in $P(uv;S)$. (Note, if $u = v$, then $\ell \ge 2$, otherwise $\ell \ge 1$.)*

*Proof.* We prove the lemma by double induction on $\ell$ and $L$, the length of the *longest* path in $P(uv;S)$. If $L = \ell$, then $\rho(uv;S) \le \delta(\ell)/p_u$, and by Lemma 8, we have $\rho(uv;S) \le \alpha^\ell$.

Assume the claim true whenever $\ell \le k$ and $L \le K$ and consider $\ell \le k+1$, $L \le K+1$.

$$\begin{aligned} \rho(uv;S) &= \frac{\alpha}{outdeg(u)} \sum_{\substack{(u,w)\in E \\ w \notin S}} \rho(wv;S), \\ &\overset{(a)}{\le} \frac{\alpha \cdot \alpha^k}{outdeg(u)} \sum_{\substack{(u,w)\in E \\ w \notin S}} 1, \\ &\le \alpha^{k+1}. \end{aligned}$$

(a) follows from the induction hypothesis because the shortest path length in $P(wv;S)$ is at most $k$ and the longest path length is at most $L$. Therefore, the claim holds for all $\ell \ge 1$ and all $L \ge \ell$. ■

### 4.1.2 Proof of Lemma 9

Consider a collection of paths $P_t$ from $v_i$ to $t$ where $t$ is the terminal node for all the paths, and does not appear as an intermediate node in any path. Let $P_t(\mathcal{S})$ be the collection of distinct prefixes. For every path $q \in P_t$, let $s(q)$ denote the terminal node of its corresponding prefix in $P_t(\mathcal{S})$. Let $\mathcal{S} = \{s_1, \ldots, s_k\}$. We can partition the paths in $P_t$ into $k$ disjoint sets $P_t^1, \ldots, P_t^k$ according to the terminal nodes of the prefixes, i.e., for every path $q \in P_t^i$, $s(q) = s_i$. Let $\Delta_{s_i}$ be the total (summed) flow of pagerank to $s_i$ along the paths in $P_t^i$.

$$I(\mathcal{S}|P_t(\mathcal{S})) = \sum_{s_i \in \mathcal{S}} \Delta_{s_i}$$

Each path in $P_t^i$ contains a suffix path from $s_i$ to $t$ in which $t$ does not appear as an intermediate node. Consider the fraction $\rho$ of $s_i$'s pagerank that flows along the distinct such suffixes to $t$. Since these suffixes are a subset of the paths in $P(s_i t;t)$, we have that $\rho \le \rho(s_i t, t)$. $I(t|P_t^i)$ can now be bounded as follows,

$$\begin{aligned} I(t|P_t^i) &= \rho \Delta_{s_i}, \\ &\le \rho(s_i t;t)\Delta_{s_i}. \end{aligned}$$

$I(t|P_t)$ is the sum of the $I(t|P_t^i)$'s, so we obtain

$$\begin{aligned} I(t|P_t) &= \sum_{i=1}^k I(t|P_t^i), \\ &\le \sum_{i=1}^k \rho(s_i t;t)\Delta_{s_i}, \\ &\overset{(a)}{\le} \alpha \sum_{i=1}^k \Delta_{s_i}, \\ &= \alpha I(\mathcal{S}|P_t(\mathcal{S})), \end{aligned}$$

where (a) follows from Lemma 19. ■

#### 4.1.3 Proof of Lemma 10

Partition the set of cycles containing $v_0$ as initial and terminal node, but not as intermediate node, into two disjoint sets $C_1$ and $C_2$. $C_1$ contains the cycles which do not contain $v_i$ and $C_2$ contains all the cycles which also contain $v_i$. (Note $C_1 = P(v_0v_0; v_0v_i)$.) Let $P_{v_0,v_i} = P(v_0v_i; v_0v_i)$ and $\rho_{v_0v_i} = \rho(v_0v_i; v_0v_i)$. (Note that $\rho_{v_0v_i} \leq \alpha$.) Every path in $C_2$ is composed of a path in $P_{v_0,v_i}$ together with a path from $v_i$ to $v_0$ in which $v_0$ appears only as a terminal node (i.e. a path in $P = P(v_iv_0; v_0)$). The fraction of $v_i$'s pagerank that flows to $v_0$ along paths in $P$ is by definition $\Delta/p_i$. Thus, the fraction of $v_0$'s pagerank that flows along cycles in $C_2$ back to $v_0$ is $\rho_{v_0v_i}\Delta/p_i$. Let $\gamma_{v_0} = \rho(v_0v_0; v_0v_i)$ be the fraction of $v_0$ pagerank that flows along cycles in $C_1$ back to $v_0$. Therefore the total fraction of $v_0$'s page rank that flows back to $v_0$ along paths in $C_1 \cup C_2$ is $\gamma_{v_0} + \rho_{v_0v_i}\Delta/p_i$. This fraction will be amplified again by the cycles in $C_1$ and $C_2$. Thus,

$$\Delta p_0^i = \Delta + \text{amp}(\Delta),$$

where amp$(x)$ satisfies

$$\text{amp}(x) = \phi x + \text{amp}(\phi x),$$

and $\phi = \phi(\Delta) = \gamma_{v_0} + \rho_{v_0v_i}\Delta/p_i < 1$. The unique solution to this equation (which can be obtained by expanding amp$(\phi x)$ repeatedly to obtain a geometric series) is

$$\text{amp}(x) = \frac{\phi x}{1-\phi}.$$

Substituting into the expression for $\Delta p_0^i$, we obtain

$$\Delta p_0^i(\Delta) = \frac{\Delta}{1 - \gamma_{v_0} - \rho_{v_0v_i}\frac{\Delta}{p_i}}.$$

To conclude, note that the right hand side is monotonically increasing in $\Delta$. ∎

#### 4.1.4 Proof of Theorem 13

Consider the attack by a single attacker $v_i$. We will show that the direct attack is best for $v_i$ independent of the rest of the graph, in particular what the other attackers do, from which the theorem will follow. For the direct individual attack not to maximize the rank (and some other attack $X$ to maximize it), there must be some vertex $u$ for which $\bar{p}_{v_0}^I \leq \bar{p}_u^I$ and $\bar{p}_{v_0}^X > \bar{p}_u^X$.

First consider the case when there are no paths from $v_0$ to $u$. Then, $p_{v_0} + \Delta p_{v_0}^I \leq p_u + \Delta p_u^I$ and $p_{v_0} + \Delta p_{v_0}^X > p_u + \Delta p_u^X$. Since $\Delta p_u^I = 0$ (no paths from $v_0$ to $u$),

$$p_u - p_{v_0} \geq \Delta p_{v_0}^I, \qquad p_u - p_{v_0} < \Delta p_{v_0}^X - \Delta p_u^X,$$

which is a contradiction because $\Delta p_{v_0}^X < \Delta p_{v_0}^I$ (Theorem 11), and $\Delta p_u^X \geq 0$.

Now consider the case when there are paths from $v_0$ to $u$. We introduce some definitions that will simplify the notation:

$$\begin{aligned}
\rho_{v_0u} &= \rho(v_0u; v_iu), \\
\rho_{v_0v_i} &= \rho(v_0v_i; v_0v_i), \\
\rho_{uv_i} &= \rho(u, v_i; v_iu), \\
\gamma_{v_0} &= \rho(v_0v_0; v_0v_i), \\
\gamma_u &= \rho(u, u; v_iu).
\end{aligned}$$

($\gamma_{v_0}$, $\rho_{v_0v_i}$ are defined as in the proof of Lemma 10.) $\gamma_{v_0}$ and $\gamma_u$ are fractions that flow along cycles.

Let $\Delta_{v_0}$ and $\Delta_u$ be the pagerank flow from $v_i$ to $v_0$ and $u$ along the respective paths $P(v_i, v_0; v_0)$ and $P(v_i, u; u)$. Then,

$$\begin{aligned} \bar{p}^I_{v_0} &= p_{v_0} + \Delta p^I_{v_0}(\Delta^I_{v_0}), \qquad & \bar{p}^I_u &= p_u + \Delta p^I_u(\Delta^I_u), \\ \bar{p}^X_{v_0} &= p_{v_0} + \Delta p^X_{v_0}(\Delta^X_{v_0}), \qquad & \bar{p}^X_u &= p_u + \Delta p^X_u(\Delta^X_u). \end{aligned}$$

($I$ denotes the direct individual attack and $X$ the other attack.) In the direct attack $I$, the only paths from $v_i$ to $u$ are through $v_0$. In the attack $X$, there may be paths from $v_i$ to $u$ that do not pass through $v_0$. Therefore, we have

$$\Delta^I_u = \rho_{v_0 u}\Delta^I_{v_0}, \qquad \Delta^X_u \geq \rho_{v_0 u}\Delta^X_{v_0}.$$

As in the proof of Lemma 10, let

$$G(x; \gamma, \rho) = \frac{x}{1 - \gamma - \rho\frac{x}{p_{v_i}}}.$$

Then,

$$\begin{aligned} \Delta p^I_{v_0}(\Delta^I_{v_0}) &= G(\Delta^I_{v_0}; \gamma_{v_0}, \rho_{v_0 v_i}), \\ \Delta p^X_{v_0}(\Delta^X_{v_0}) &= G(\Delta^X_{v_0}; \gamma_{v_0}, \rho_{v_0 v_i}), \end{aligned}$$

and,

$$\begin{aligned} \Delta p^I_u(\Delta^I_u) &= G(\Delta^I_u; \gamma_u, \rho_{u v_i}), \\ &= G(\rho_{v_0 u}\Delta^I_{v_0}; \gamma_u, \rho_{u v_i}), \\ &= \rho_{v_0 u} G(\Delta^I_{v_0}; \gamma_u, \rho_{v_0 u}\rho_{u v_i}), \\ \Delta p^X_u(\Delta^X_u) &= G(\Delta^X_u; \gamma_u, \rho_{u v_i}), \\ &\overset{(a)}{\geq} G(\rho_{v_0 u}\Delta^X_{v_0}; \gamma_u, \rho_{u v_i}), \\ &= \rho_{v_0 u} G(\Delta^X_{v_0}; \gamma_u, \rho_{v_0 u}\rho_{u v_i}). \end{aligned}$$

(a) follows because $G$ is monotonic in $x$, and we have used the identity $G(\lambda x; \gamma, \rho) = \lambda G(x; \gamma, \lambda\rho)$. $u$ is such that $\bar{p}^I_u - \bar{p}^I_{v_0} \geq 0$ and $\bar{p}^X_u - \bar{p}^X_{v_0} < 0$. Thus,

$$\begin{aligned} p_u - p_{v_0} &\geq G(\Delta^I_{v_0}; \gamma_{v_0}, \rho_{v_0 v_i}) - \rho_{v_0 u} G(\Delta^I_{v_0}; \gamma_u, \rho_{v_0 u}\rho_{u v_i}), \\ p_u - p_{v_0} &< G(\Delta^X_{v_0}; \gamma_{v_0}, \rho_{v_0 v_i}) - \rho_{v_0 u} G(\Delta^X_{v_0}; \gamma_u, \rho_{v_0 u}\rho_{u v_i}). \end{aligned}$$

Combining these two equations, we find that

$$F(\Delta^X_{v_0}, \Delta^I_{v_0}; \gamma_{v_0}, \rho_{v_0 v_i}) > \rho_{v_0 u} F(\Delta^X_{v_0}, \Delta^I_{v_0}; \gamma_u, \rho_{v_0 u}\rho_{u v_i}),$$

where

$$\begin{aligned} F(x_1, x_2; \gamma, \rho) &= G(x_1; \gamma, \rho) - G(x_2; \gamma, \rho), \\ &= \frac{(x_1 - x_2)(1 - \gamma)}{(1 - \gamma - \rho\frac{x_1}{p_{v_i}})(1 - \gamma - \rho\frac{x_2}{p_{v_i}})} \end{aligned}$$

Since $\Delta^X_{v_0} < \Delta^I_{v_0}$ (Theorem 11), we obtain

$$\frac{\rho_{v_0 u} F(\Delta^X_{v_0}, \Delta^I_{v_0}; \gamma_u, \rho_{v_0 u}\rho_{u v_i})}{\Delta^X_{v_0} - \Delta^I_{v_0}} > \frac{F(\Delta^X_{v_0}, \Delta^I_{v_0}; \gamma_{v_0}, \rho_{v_0 v_i})}{\Delta^X_{v_0} - \Delta^I_{v_0}}$$

Let $\rho_{00} = \rho(v_0 v_0; v_0 v_i u)$, and let $\rho_{uu} = \rho(uu; v_0 v_i u)$. Let $Q = (1 - \rho_{00})(1 - \rho_{uu})$. We will need the following lemmas to complete the proof. We will prove the lemmas after the proof of the theorem.

**Lemma 20.** $1-\gamma_{v_0} = Q\cdot(1-\gamma_u)$.

**Lemma 21.** $\rho_{v_0v_i} \geq Q\rho_{v_0u}\rho_{uv_i}$.

By Lemma 21 and the monotonicity of $F$ with respect to $\rho$, we have

$$\frac{\rho_{v_0u}F(\Delta^X_{v_0},\Delta^I_{v_0};\gamma_u,\rho_{v_0u}\rho_{uv_i})}{\Delta^X_{v_0}-\Delta^I_{v_0}} > \frac{F(\Delta^X_{v_0},\Delta^I_{v_0};\gamma_{v_0},Q\rho_{v_0u}\rho_{uv_i})}{\Delta^X_{v_0}-\Delta^I_{v_0}}$$

or that,

$$\begin{aligned}
& \frac{\rho_{v_0u}(1-\gamma_u)}{(1-\gamma_u-\rho_{v_0u}\rho_{uv_i}\frac{\Delta^X_{v_0}}{p_{v_i}})(1-\gamma_u-\rho_{v_0u}\rho_{uv_i}\frac{\Delta^I_{v_0}}{p_{v_i}})} \\
> \quad & \frac{(1-\gamma_{v_0})}{(1-\gamma_{v_0}-Q\rho_{v_0u}\rho_{uv_i}\frac{\Delta^X_{v_0}}{p_{v_i}})(1-\gamma_{v_0}-Q\rho_{v_0u}\rho_{uv_i}\frac{\Delta^I_{v_0}}{p_{v_i}})}, \\
\stackrel{(a)}{=} \quad & \frac{Q(1-\gamma_u)}{Q^2(1-\gamma_u-\rho_{v_0u}\rho_{uv_i}\frac{\Delta^X_{v_0}}{p_{v_i}})(1-\gamma_u-\rho_{v_0u}\rho_{uv_i}\frac{\Delta^I_{v_0}}{p_{v_i}})},
\end{aligned}$$

where (a) follows using Lemma 20. After some algebraic manipulations, we obtain

$$Q\rho_{v_0u} = \frac{1-\rho_{00}}{1-\rho_{uu}}\rho_{v_0u} > 1.$$

In any attack, $v_0$'s pagerank flows to $u$ with attenuation $\rho(v_0u;v_0u)$ amplified by $1/(1-\rho(uu;v_0))$. Since $p_u$'s pagerank cannot be smaller that what flows from $v_0$, we have

$$\begin{aligned}
p_u \quad & \geq \quad \frac{\rho(v_0u;v_0u)}{1-\rho(uu;v_0)}p_{v_0}, \\
& \stackrel{(a)}{\geq} \quad \frac{\rho(v_0u;v_0v_iu)}{1-\rho(uu;v_0v_iu)}p_{v_0}, \\
& \stackrel{(b)}{=} \quad \frac{(1-\rho(v_0v_0;v_0v_iu))\rho(v_0u;v_iu)}{1-\rho(uu;v_0v_iu)}p_{v_0}, \\
& \stackrel{(c)}{=} \quad \frac{(1-\rho_{00})\rho_{v_0u}}{1-\rho_{uu}}p_{v_0}, \\
& > \quad p_{v_0}.
\end{aligned}$$

(a) follows from Lemma 17; (b) follows because using Lemma 18,

$$\rho(v_0u;v_iu) = \frac{\rho(v_0u;v_0v_iu)}{1-\rho(v_0v_0;v_0v_iu)};$$

and, (c) follows from the definitions of $\rho_{00},\rho_{uu},\rho_{v_0u}$. Thus, $p_u > p_{v_0}$ for *any* attack, in particular, for the attack $X$, which contradicts the fact that $\bar{p}^X_{v_0} > \bar{p}^X_u$. This contradiction implies that no such vertex $u$ can exist, which concludes the proof of the theorem. ∎

### 4.1.5 Proof of Lemma 20

We use the same notation as in the proof of Theorem 13. Let $S=\{v_0,v_i,u\}$. $\gamma_0=\rho(v_0v_0;v_0v_i)$ is the fraction of $v_0$'s rank flow back to $v_0$ along paths in $P(v_0v_0;v_0v_i)$. The paths in $P(v_0v_0;v_0v_i)$ can be partitioned into paths that contain $u$ and paths that do not. The paths that contain $u$ are paths in $P(v_0u;S)$ concatenated with paths in $P(uu;S)$ concatenated with paths in $P(uv_0;S)$. Therefore, using Lemma 18,

$$\gamma_0 = \rho(v_0v_0;S) + \frac{\rho(v_0u;S)\rho(uv_0;S)}{1-\rho(uu;S)}.$$

Applying similar reasoning to $\gamma_u$, and using the definitions for $\rho_{00}, \rho_{uu}$, we obtain

$$\begin{aligned} \gamma_0 &= \rho_{00} + \frac{\rho(v_0u;S)\rho(uv_0;S)}{1-\rho_{uu}}, \\ \gamma_u &= \rho_{uu} + \frac{\rho(v_0u;S)\rho(uv_0;S)}{1-\rho_{00}}. \end{aligned}$$

Let $A = \rho(v_0u;S)\rho(uv_0;S)$. We find that

$$\begin{aligned} 1-\gamma_0 &= \frac{(1-\rho_{00})(1-\rho_{uu})-A}{1-\rho_{uu}}, \\ 1-\gamma_u &= \frac{(1-\rho_{00})(1-\rho_{uu})-A}{1-\rho_{00}}, \end{aligned}$$

It now follows that $(1-\gamma_0) = Q(1-\gamma_u)$. ∎

#### 4.1.6 Proof of Lemma 21

We use the same notation as in the proof of Theorem 13. Let $S = \{v_0, v_i, u\}$. Then,

$$\begin{aligned} \rho_{v_0v_i} &= \rho(v_0v_i;S) + \frac{\rho(v_0u;S)\rho(uv_i;S)}{1-\rho_{uu}}, \\ \rho_{v_0u} &= \frac{\rho(v_0u;S)}{1-\rho_{00}}, \\ \rho_{uv_i} &= \rho(uv_i;S) + \frac{\rho(uv_0;S)\rho(v_0v_i;S)}{1-\rho_{00}}. \end{aligned}$$

Therefore, we find that

$$\begin{aligned} Q\rho_{v_0u}\rho_{uv_i} &= \frac{\rho(v_0u;S)\rho(uv_i;S)}{1-\rho_{uu}} + \frac{\rho(v_0u;S)\rho(uv_0;S)\rho(v_0v_i;S)}{(1-\rho_{uu})(1-\rho_{00})}, \\ &= \rho_{v_0v_i} - \rho(v_0v_i;S) + \frac{\rho(v_0u;S)\rho(uv_0;S)\rho(v_0v_i;S)}{(1-\rho_{uu})(1-\rho_{00})}. \end{aligned}$$

After rearranging terms, we obtain

$$\begin{aligned} \rho_{v_0v_i} - Q\rho_{v_0u}\rho_{uv_i} &= \rho(v_0v_i;S) \cdot \left(1 - \frac{\rho(v_0u;S)}{1-\rho_{uu}} \cdot \frac{\rho(uv_0;S)}{1-\rho_{00}}\right), \\ &\overset{(a)}{=} \rho(v_0v_i;S) \cdot \left(1 - \rho(v_0u;v_iu)\rho(uv_0;v_0v_i)\right). \end{aligned}$$

(a) follows from Lemma 18. To conclude, note that $\rho(v_0u;v_iu)\rho(uv_0;v_0v_i) \leq \alpha^2$ (Lemma 19), and so the right hand side is $\geq 0$. ∎

### 4.2 The Optimal Disguised Attack

For the optimal disguised attack, every path from an attacker to the victim must have length $\geq \ell$. We only consider the case that such attacks are possible, in particular, $U_{\ell-1}$ is not empty.

Consider the graph with the edge set $E_u = E \cup (v_i, u)$. Let $P_u(vw; x_1, \ldots, x_k)$ and $\rho_u(vw; x_1, \ldots, x_k)$ be defined with respect to the edge set $E_u$ in exactly the same way that $P(vw; x_1, \ldots, x_k)$ and $\rho(vw; x_1, \ldots, x_k)$ were defined in the previous section. Note that

$$V_i(u) = f_u(v_i; v_0) = \rho_u(v_i v_0; v_0),$$

and more generally,

$$f_u(v, w) = \rho_u(vw; w).$$

Let $\rho_{max}(v)$ be the maximum forward rank (with respect to $v_0$) of any 1-neighbor of $v$,

$$\rho_{max}(v) = \max_{(v,w)\in E_u} \rho_u(wv_0; v_0)$$

From the forward equation, by replacing the summand by the largest term, we have,

**Lemma 22.** *$\rho_u(vv_0; v_0) \leq \alpha\rho_{max}(v)$, with equality* iff *$\rho_u(w_1 v_0; v_0) = \rho_u(w_2 v_0; v_0)$ for all $w_1, w_2$ such that $(u, w_1), (u, w_2) \in E_u$.*

**Lemma 23.** *There is at least one vertex $w^* \in U_{\ell-1}$ with $\rho_u(uv_0; v_0) \leq \rho_u(w^* v_0; v_0)$*

*Proof.* Consider $\rho_u(uv_0; v_0)$. We can assume that $\rho_u(uv_0; v_0) > 0$ (i.e., $d(u, v_0) < \infty$), and that $d(u, v_0) > \ell - 1$, as otherwise there is nothing to prove. Choose $w_1$ in the 1-neighborhood of $u$ such that $\rho_u(wv_0; v_0) = \rho_{max}(u)$. If there is more than one possibile choice for $w_1$, select the choice for which $d(w_1, v_0)$ is minimized, breaking any further ties arbitrarily. If $d(w_1, v_0) = \ell - 1$ then we stop, otherwise we define $w_2$ in a similar way to $w_1$: $w_2$ is a vertex in $N_1(w_1)$ such that $\rho_u(w_2 v_0; v_0) = \rho_{max}(w_1)$. In general, $w_{i+1} = \underset{(w_i, w_{i+1})\in E_u}{\text{argmax}} \rho_u(w_{i+1} v_0; v_0)$, breaking ties according to distance. By Lemma 22, since $\alpha \leq 1$, for the sequence $u, w_1, w_2, \ldots$,

$$\rho_u(uv_0; v_0) \leq \rho_u(w_1 v_0; v_0) \leq \rho_u(w_2 v_0; v_0) \leq \cdots.$$

Further, if $\rho_u(w_i v_0; v_0) = \rho_u(w_{i+1} v_0; v_0)$ (which can only happen if $\alpha = 1$ and all neighbors have the same $\rho$), then, since the ties were broken by distance, $d(w_i, v_0) > d(w_{i+1}, v_0)$. Thus, there are no repetitions in the sequence $u, w_1, w_2, \ldots$. Since there is a path from $u$ to $v_0$ and $d(u, v_0) > \ell - 1$, by the pigeon hole principle, we conclude that at least one vertex $w^*$ in this sequence is distance $\ell - 1$ from $v_0$, ■

Note that equality in the Lemma can only occur if $\alpha = 1$, thus for $\alpha < 1$, it is strictly better to be in $U_{\ell-1}$ than not. To prove Lemma 14, we will show that $V_i(w^*) \geq V_i(u)$.

### 4.2.1 Proof of Lemma 14

Suppose that the maximum is attained for a vertex $u$ with $d(u, v_0) > \ell - 1$. Let $w^* \in U_{\ell-1}$ be such that $\rho_u(w^* v_0; v_0) \geq \rho_u(uv_0; v_0)$ (Lemma 23 guarantees the existence of such a vertex). We show that $V_i(u) \leq V_i(w^*)$. From the definitions of $V_i(u)$ and $V_i(w^*)$, we have that

$$\begin{aligned}
V_i(u) &= \rho_u(v_i v_0; v_0), \\
&= \alpha\rho_u(uv_0; v_0), \\
&\overset{(a)}{\leq} \alpha\rho_u(w^* v_0; v_0); \\
V_i(w^*) &= \rho_{w^*}(v_i v_0), \\
&= \alpha\rho_{w^*}(w^* v_0; v_0).
\end{aligned}$$

(a) follows from the definition of $w^*$. By considering the paths which reuse $v_i$ and those which do not, we have that

$$\begin{aligned}
\rho_u(w^*v_0; v_0) &= \rho_u(w^*v_0; v_iv_0) + \rho_u(w^*v_i; v_iv_0)\rho_u(v_iv_0; v_0), \\
&\overset{(a)}{=} \rho_u(w^*v_0; v_iv_0) + \alpha\rho_u(w^*v_i; v_iv_0)\rho_u(uv_0; v_0), \\
&\overset{(b)}{\leq} \rho_u(w^*v_0; v_iv_0) + \alpha\rho_u(w^*v_i; v_iv_0)\rho_u(w^*v_0; v_0), \\
&\overset{(c)}{=} \rho_{w^*}(w^*v_0; v_iv_0) + \alpha\rho_{w^*}(w^*v_i; v_iv_0)\rho_u(w^*v_0; v_0); \\
\rho_{w^*}(w^*v_0; v_0) &= \rho_{w^*}(w^*v_0; v_iv_0) + \rho_{w^*}(w^*v_i; v_iv_0)\rho_{w^*}(v_iv_0; v_0), \\
&\overset{(d)}{=} \rho_{w^*}(w^*v_0; v_iv_0) + \alpha\rho_{w^*}(w^*v_i; v_iv_0)\rho_{w^*}(w^*v_0; v_0).
\end{aligned}$$

(a) follows because the only edge from $v_i$ in $E_u$ is $(v_i, u)$, and similarly for (d). (b) follows from the definition of $w^*$. (c) follows because only difference between $E_u$ and $E_{w^*}$ is that the edge $(v_i, u)$ in $E_u$ is replaced by the edge $(v_i, w^*)$ in $E_{w^*}$. Therefore all paths that do not include $v_i$ as an intermediate node are identical in $E_u$ and $E_w$, and so the corresponding $\rho$'s are equal. Since $\rho_{w^*}(w^*v_0; v_iv_0) > 0$, $\rho_{w^*}(w^*v_i; v_iv_0) < 1$, solving for $\rho_{w^*}(w^*v_0; v_0)$ and $\rho_u(w^*v_0; v_0)$, we get

$$\begin{aligned}
\rho_{w^*}(w^*v_0; v_0) &= \frac{\rho_{w^*}(w^*v_0; v_iv_0)}{1 - \alpha\rho_{w^*}(w^*v_i; v_iv_0)}, \\
\rho_u(w^*v_0; v_0) &\leq \frac{\rho_{w^*}(w^*v_0; v_iv_0)}{1 - \alpha\rho_{w^*}(w^*v_i; v_iv_0)}, \\
&= \rho_{w^*}(w^*v_0; v_0).
\end{aligned}$$

Thus, $V_i(u) \leq V_i(w^*)$. ∎

### 4.2.2 Proof of Theorem 15

Lets consider an arbitrary attack $X$ in which $v_i$ has links to $w_1, w_2, \ldots, w_m$, where $d(w_j, v_0) \geq \ell - 1$ for $j \in [1, m]$. Suppose that $v_i$ has $k_j$ links to $w_j$. Let $E_X = E \cup \{(v_i, w_j)_{k_j}\}_{j=1}^m$ be the augmented edge set, where $(v_i, w_j)_{k_j}$ represents $k_j$ copies of $(v_i, w_j)$. Let $\rho_X(v_iv_0; v_0)$ be the fraction of $v_i$'s rank that flows to $v_0$ along paths in $P_X(v_iv_0; v_0)$,

$$\Delta_{v_0}^X = \rho_X(v_iv_0; v_0)p_{v_i}.$$

For the single attack $I$, $v_i$ has only one link to $w^* \in U_{\ell-1}$, where $w^*$ is such that $V_i(w^*) \geq V_i(u)$ for all $u$ such that $d(u, v_0) \geq \ell - 1$.

$$\Delta_{v_0}^I = \rho_{w^*}(v_iv_0; v_0)p_{v_i}.$$

By Lemma 10, it suffices to show that $\Delta^I_{v_0} \geq \Delta^X_{v_0}$, i.e. that $\rho_{w^*}(v_iv_0; v_0) \geq \rho_X(v_iv_0; v_0)$. Let $K = \sum_{j=1}^m k_j$. Using (3),

$$\begin{aligned}
\rho_X(v_iv_0; v_0) &= \frac{\alpha}{K}\sum_{j=1}^m k_j \rho_X(w_jv_0; v_0),\\
&\stackrel{(a)}{=} \frac{\alpha}{K}\sum_{j=1}^m k_j \left[\rho_X(w_jv_0; v_iv_0) + \rho_X(w_jv_i; v_iv_0)\rho_X(v_iv_0; v_0)\right],\\
&\stackrel{(b)}{=} \frac{\alpha}{K}\sum_{j=1}^m k_j \left[\rho_{w_j}(w_jv_0; v_iv_0) + \rho_{w_j}(w_jv_i; v_iv_0)\rho_X(v_iv_0; v_0)\right],\\
&\stackrel{(c)}{=} \frac{\frac{\alpha}{K}\sum_{j=1}^m k_j\rho_{w_j}(w_jv_0; v_iv_0)}{1 - \frac{\alpha}{K}\sum_{j=1}^m k_j\rho_{w_j}(w_jv_i; v_iv_0)},\\
&\leq \frac{\alpha\rho_{\bar{w}}(\bar{w}v_0; v_iv_0)}{1 - \alpha\rho_{\bar{w}}(\bar{w}v_i; v_iv_0)},\\
&\stackrel{(d)}{=} \rho_{\bar{w}}(v_iv_0; v_0),
\end{aligned}$$

where $\bar{w} = \text{argmax}_{w_j} \rho_{w_j}(w_jv_0; v_iv_0)$. (a) follows by partitioning the paths from $w_j$ to $v_0$ into those that use $v_i$ and those that do not; (b) follows because the paths that do not use $v_i$ are identical in both $E_X$ and $E_{w_j}$; (c) follows after solving for $\rho_X(v_iv_0; v_0)$; and (d) follows after solving for $\rho_{w_j}(v_iv_0; v_0)$ in

$$\rho_{w_j}(v_iv_0; v_0) = \alpha\left[\rho_{w_j}(w_jv_0; v_iv_0) + \rho_{w_j}(v_i; v_iv_0)\rho_{w_j}(v_iv_0; v_0)\right].$$

To conclude, note that by the definition of $w^*$, $\rho_{w^*}(v_iv_0; v_0) \geq \rho_{\bar{w}}(v_iv_0; v_0)$. ∎

#### 4.2.3 Proof of Theorem 16

Let $X$ be an optimal attack in which each attacker's only link is to a node in $U_{\ell-1}$ (not necessarily the same node for each attacker). By Theorem 15, such an optimal attack exists. Suppose that attacker $v_i$ points to node $w_i \in U_{\ell-1}$. Then,

$$\rho(v_iv_0; v_0) = \alpha\rho(w_iv_0; v_0).$$

Let $A = \{v_0, v_1, \dots, v_K\}$ denote the set containing the attackers and the victim. We use the notation $\rho_{uv} = \rho(uv; A)$ to be the fraction of rank flowing from $u$ to $v$ along paths that do not contain a node of $A$ as an intermediate node. Let $v^*$ be the attacker satisfying

$$\rho(v^*v_0; v_0) \geq \rho(v_iv_0; v_0),$$

where $v^*$ and $v_i$ are attackers, and denote the node that $v^*$ points to as $w^*$. Then, $\rho(w^*v_0; v_0) \geq \rho(w_iv_0; v_0)$ for all $i \in [1, K]$.

We now consider the attack $\bar{X}$ in which every attacker only points to $w^*$. We will show that for every node $v$, $\bar{\rho}(vv_0; v_0) \geq \rho(vv_0; v_0)$, where $\rho$ (resp. $\bar{\rho}$) is the fraction of $v$'s rank that propagates to $v_0$ under attack $X$ (resp. $\bar{X}$). First consider $w^*$. We have

$$\begin{aligned}
\rho(w^*v_0; v_0) &= \rho_{w^*v_0} + \alpha\sum_{i=1}^K \rho_{w^*v_i}\rho(w_iv_0; v_0),\\
&\stackrel{(a)}{\leq} \rho_{w^*v_0} + \alpha\sum_{i=1}^K \rho_{w^*v_i}\rho(w^*v_0; v_0),\\
&\stackrel{(b)}{\leq} \frac{\rho_{w^*v_0}}{1 - \alpha\sum_{i=1}^K \rho_{w^*v_i}}.
\end{aligned}$$

(a) follows because $\rho(w_i v_0; v_0) \le \rho(w^* v_0; v_0)$, and (b) after solving for $\rho(w^* v_0; v_0)$. Paths that do not pass through an attacker are identical in the attack $\bar{X}$ and $X$. Thus, $\bar{\rho}_{uv} = \rho_{uv}$ and so

$$\begin{aligned}
\bar{\rho}(w^* v_0; v_0) &\overset{(a)}{=} \rho_{w^* v_0} + \alpha \sum_{i=1}^{K} \rho_{w^* v_i} \bar{\rho}(w^* v_0; v_0), \\
&\overset{(b)}{=} \frac{\rho_{w^* v_0}}{1 - \alpha \sum_{i=1}^{K} \rho_{w^* v_i}}.
\end{aligned}$$

(a) follows because every attacker in $\bar{X}$ points to $w^*$ and (b) after solving for $\bar{\rho}(w^* v_0; v_0)$. Thus we conclude that $\rho(w^* v_0; v_0) \le \bar{\rho}(w^* v_0; v_0)$. Now consider an arbitrary node $v$.

$$\begin{aligned}
\rho(v v_0; v_0) &= \rho_{v v_0} + \sum_{i=1}^{K} \rho_{v v_i} \rho(v_i v_0; v_0), \\
&= \rho_{v v_0} + \alpha \sum_{i=1}^{K} \rho_{v v_i} \rho(w_i v_0; v_0), \\
&\overset{(a)}{\le} \rho_{v v_0} + \alpha \sum_{i=1}^{K} \rho_{v v_i} \rho(w^* v_0; v_0), \\
&\overset{(b)}{\le} \rho_{v v_0} + \alpha \sum_{i=1}^{K} \rho_{v v_i} \bar{\rho}(w^* v_0; v_0), \\
&= \bar{\rho}(v v_0; v_0).
\end{aligned}$$

(a) follows by definition of $w^*$, and (b) because $\rho(w^* v_0; v_0) \le \bar{\rho}(w^* v_0; v_0)$, thus the magnitude of $\bar{X}$ is at least as large as the magnitude of $X$. ■

# 5 Experimental results

In this section, we give some preliminary experimental results that quantify the effectiveness of link bombs in various environments. There are four main degrees of freedom we explore: the nature of the graph, including its connectivity or edge density; the prominence (pagerank) of the attackers; the prominence of the victim; and, the value of $\alpha$.

We ran our experiments on three types of graphs: *Random* is an Erdös-Reyni type ($G(n, p)$) random graph with edge probability $p$; $BA$ (Barabási-Albert) is a preferential-attachment random graph with 5 outgoing edges per vertex [3]; (Such graphs are known to have power-law in-degree distributions, and since we add the vertices sequentially, there are no cycles.) $MWDTA$ is a modified "Winner's don't take all" random graph in which every node has at least one out-going edge [19]. (Such graphs are known to model certain characteristics of the world wide web graph such as power-law in and out-degree distributions.). The main difference between $MWDTA$ and $BA$ random graphs is that in $MWDTA$, a larger number of nodes will have significant in-degree, whereas in $BA$ a few nodes have very large in-degrees. In order to make fair comparisons, we normalize graphs from different random graph models (*Random*, $BA$ or $MWDTA$) to have the same expected number of edges.

First, we generate a random graph with 1,000 nodes, and randomly select 10 attackers and a victim. We then remove outgoing edges from the attackers and perform a pagerank computation, obtaining:

$p_0$: the page rank of the victim;
$p_A$: the average pagerank of the attackers;
$f_p(p)$: the pagerank distribution in the graph;
$\sigma_p$: the std. dev. of the pagerank distribution.

We only show results for two of the attacks described in Section 3.1: the optimal direct individual attack $I$, and the cycle attack $C$ (the results for other suboptimal attacks are similar). Each attack is repeated a number of times on randomly generated graphs to increase the statistical significance of the results. We use the following measures of success for attack $X$,

$$\begin{aligned} G(X) &= \text{Gain} = \frac{\Delta p_0^X}{p_0}, \\ \bar{G}(X) &= \text{Normalized Gain} = \frac{\Delta p_0^X}{\sigma_p}, \\ D(X) &= \text{Discrepancy Factor} = \frac{G(I)}{G(X)}, \\ \bar{D}(X) &= \text{Normalized Discrepancy} = \bar{G}(I) - \bar{G}(X). \end{aligned}$$

The pagerank distribution $f_p(p)$ generally affects the effectiveness of an attack. Figure 1 shows pagerank distributions for the various random graphs. As can be seen, *Random* has a (near) Normal distribution, compared with $BA$ and $MWDTA$ which have power-law type distributions in which $MWDTA$ appears to have a slightly fatter tail than $BA$.

Some detailed results on the effectiveness of the attacks are shown in Figure 2: (a) shows how connectivity (number of edges) in *Random* graphs with different $p$ affects the attack; (b) shows different graph types; (a) and (b) show the dependence on the prominence of the attackers, and (c) shows the dependence on the prominence of the victim; (f) shows the dependence on $\alpha$. Figure 3 shows some results for the *rank* (as opposed to the pagerank). We give a summary of the results below.

*Higher Density:* All attacks decrease in magnitude (new edges have little additional effect when the graph is already dense).

*Graph type:* Prominence of attackers has (by far) the largest impact in *Random* graphs, as compared to $BA$ and $MWDTA$. (Pageranks in *Random* graphs are “concentrated” around the mean, so any bias in the victim’s pagerank results in it becoming extreme. This is less so for $BA$ and even less so for $MWDTA$.).

*Higher Prominence of Attackers:* Stronger attack.

*Higher Prominence of Victim:* Attacks become less effective and $D(C)$ decreases (diminishing returns).

*Lower $\alpha$:* $D(C)$ increases (it is more costly to divert from the individual attack).

*Rank:* For random graphs, an attack usually results in a top ranking for the victim, which is not usually the case for $BA$ and $MWDTA$ graphs.

## 6 Discussion

We have shown that the best attack is the direct individual attack, in particular: *any* organized structure among the attackers reduces the impact of the attack; links that cycle back to attackers in an attempt to boost their pageranks are detrimental. The discrepancy between the optimal individual attack and suboptimal attacks can strongly depend on the graph type through the initial pagerank distribution. Our results indicate conditions that offer resistance to rank manipulation: dense, power-low type graphs in which victims already have high rank, attackers have low rank

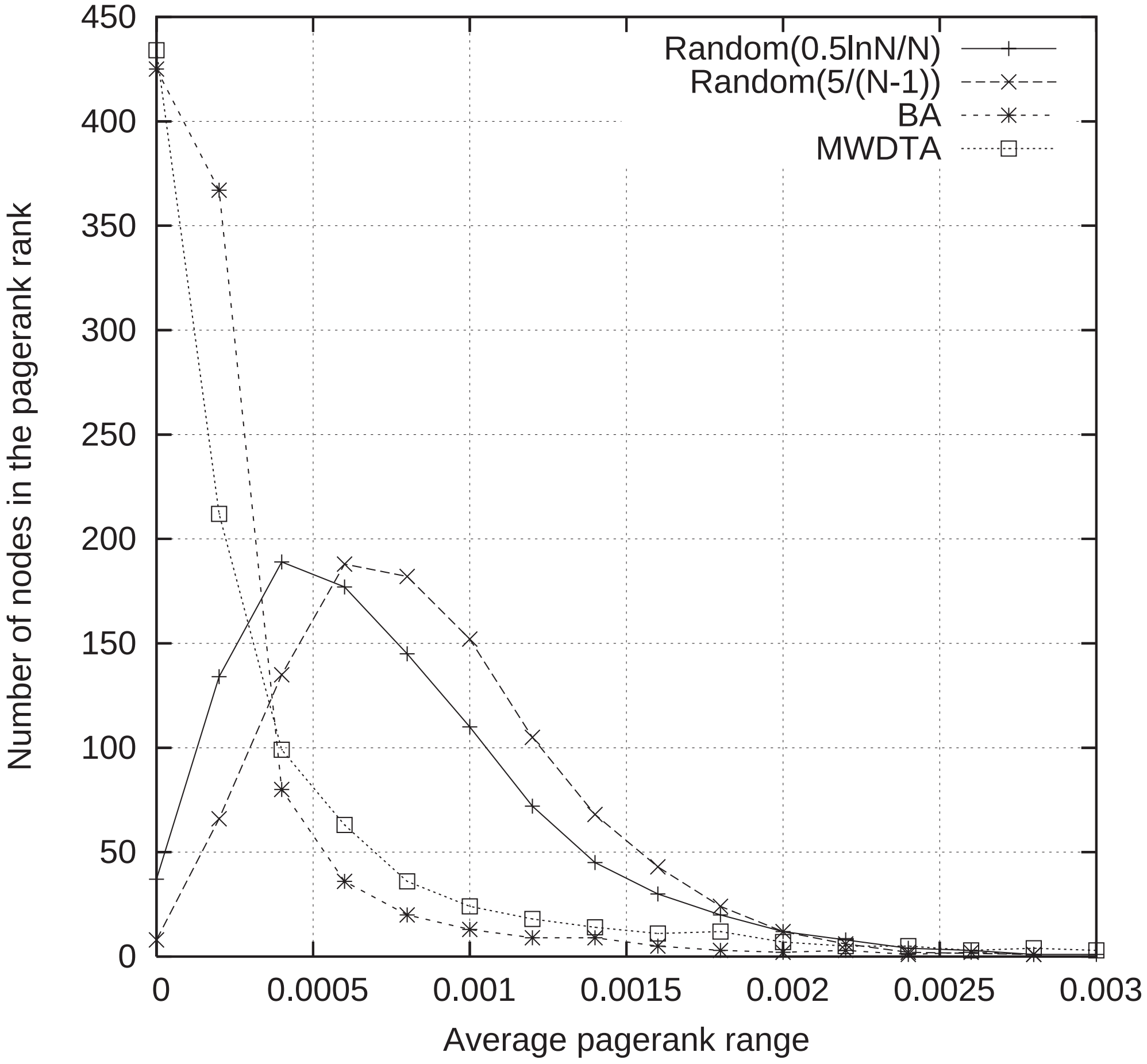

Figure 1: Pagerank distributions of different graphs

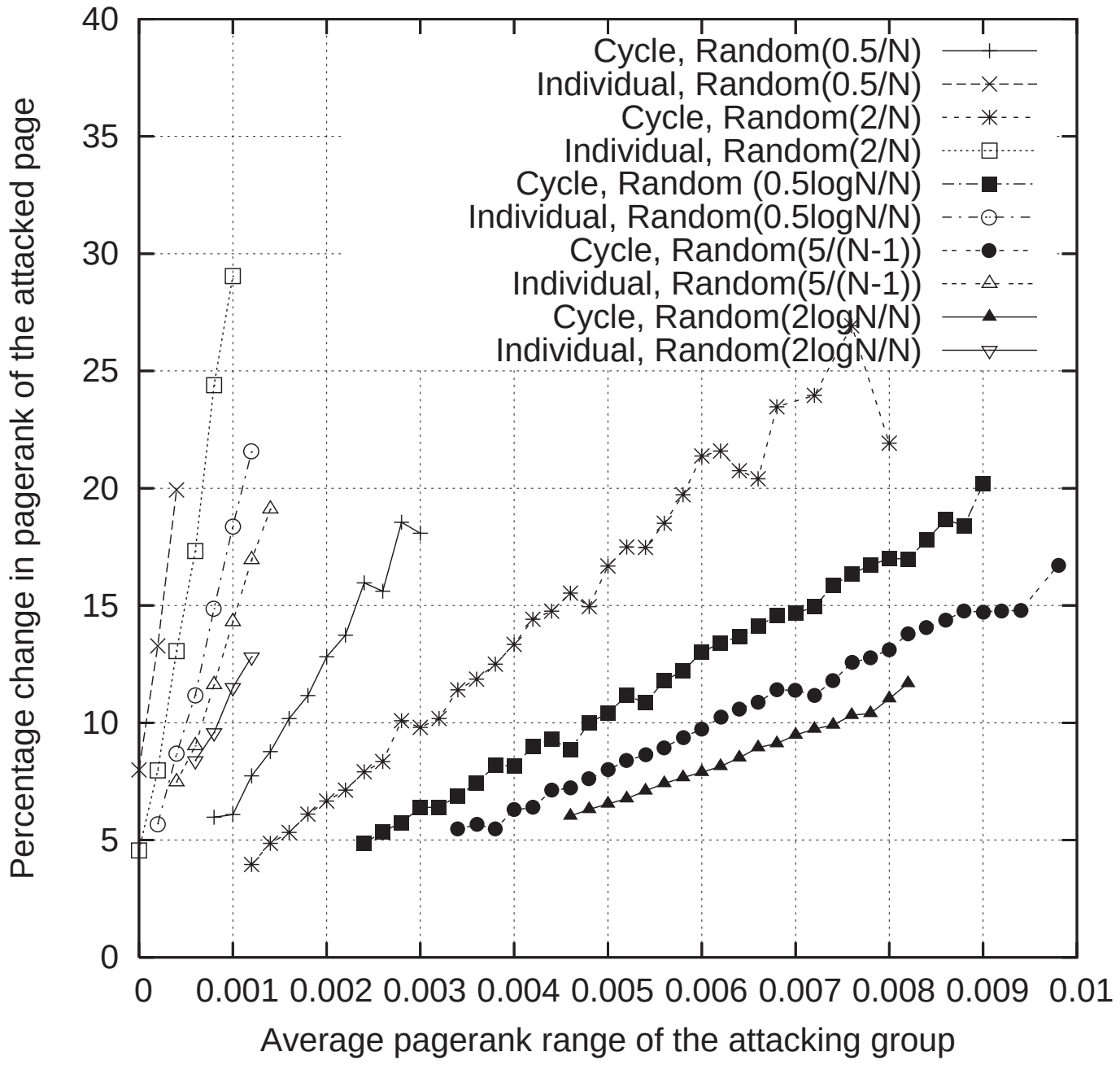

(a) Graphs with different edge densities.

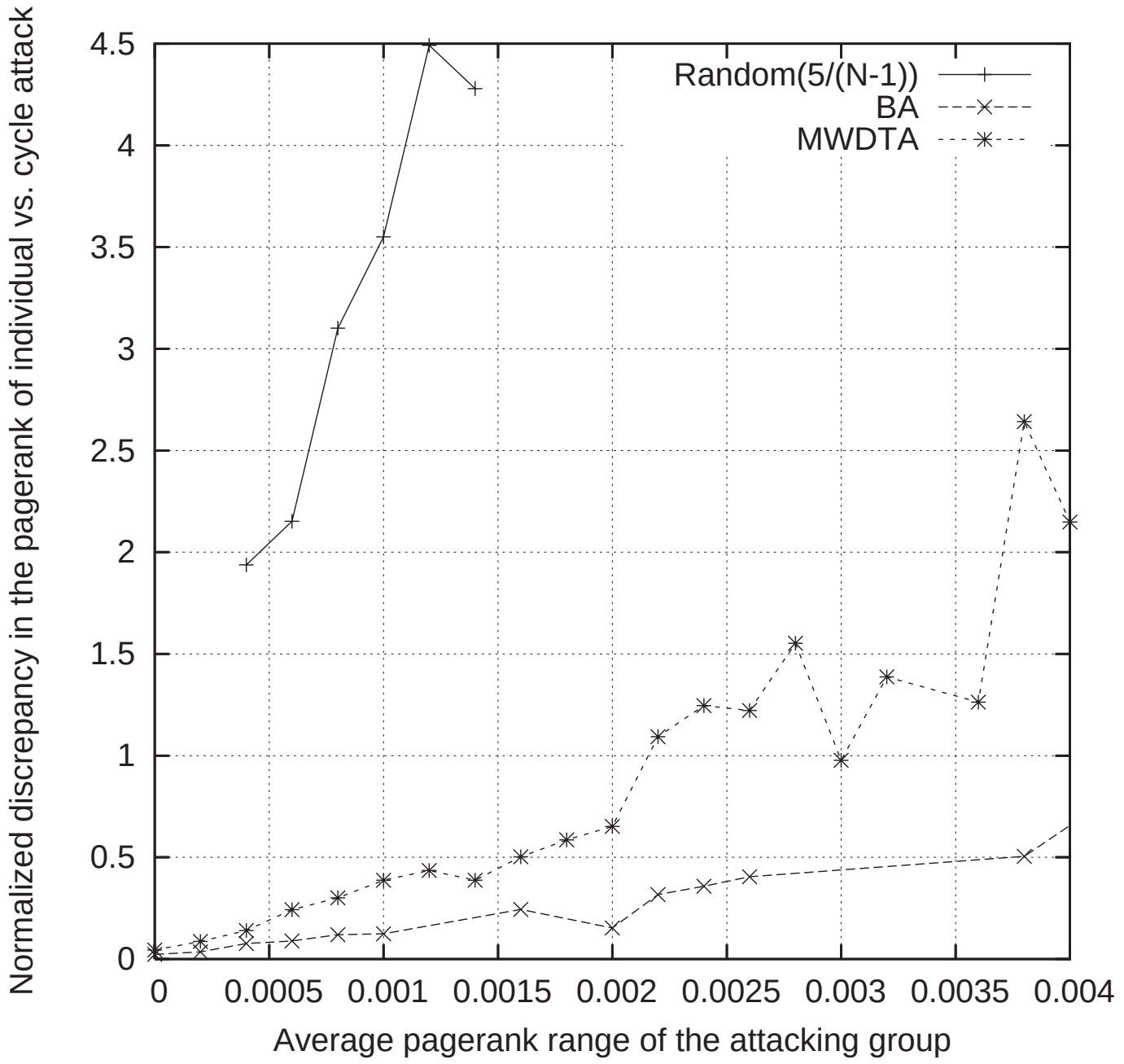

(b) Attacks in different graph types

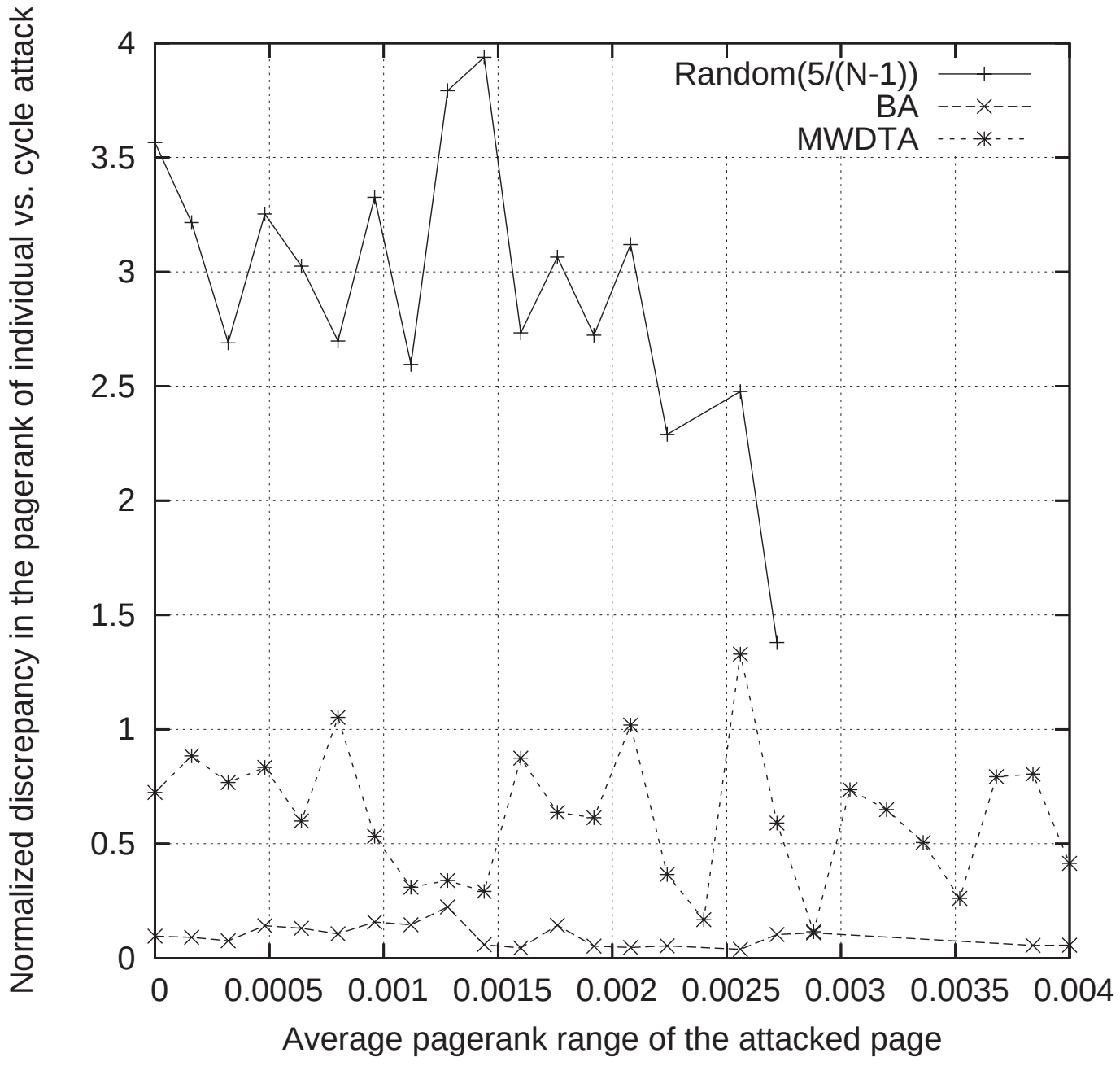

(c) Dependence on victim's pagerank

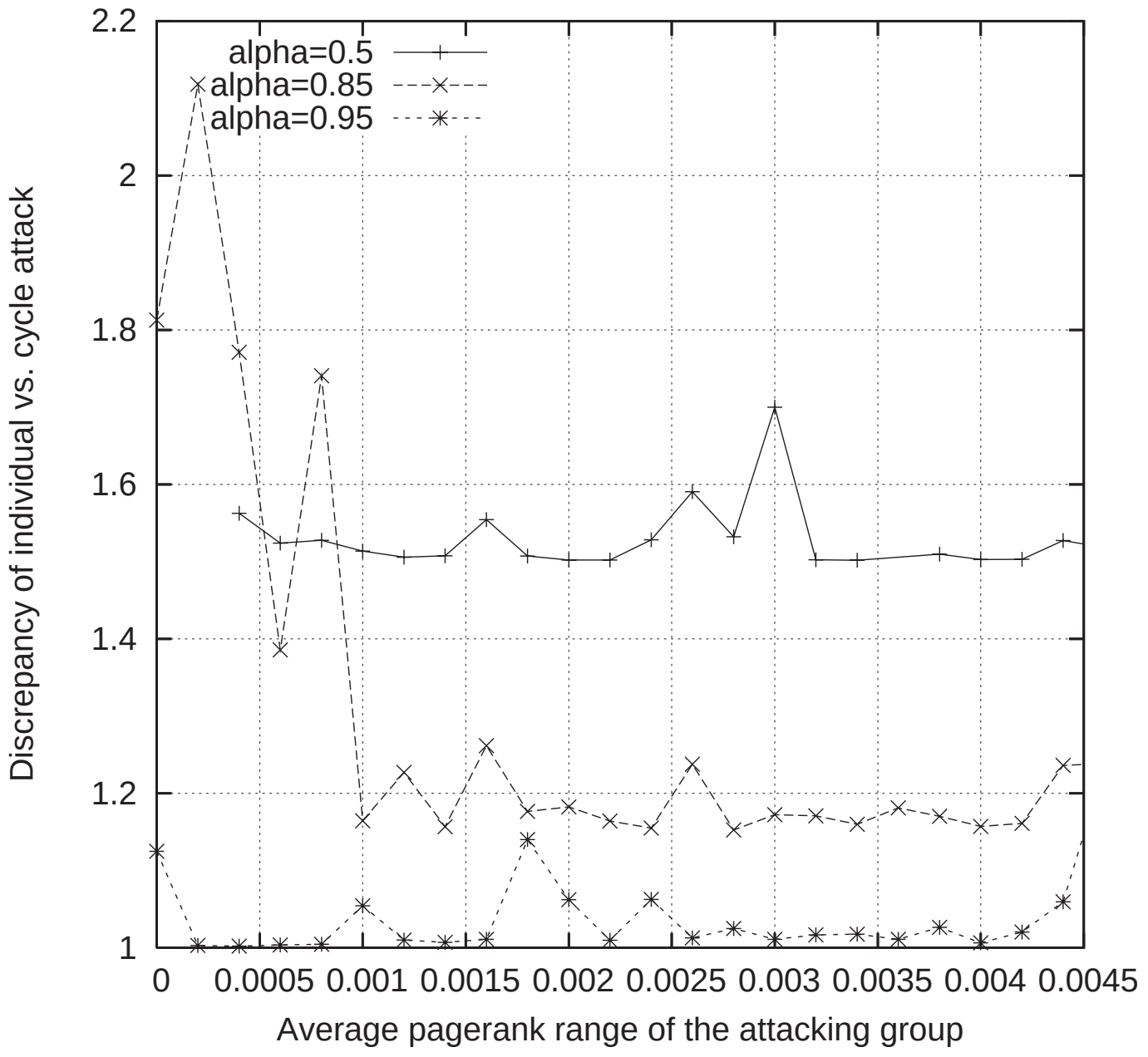

(d) Dependence on $\alpha$ (MWDTA)

Figure 2: Experimental results on pagerank

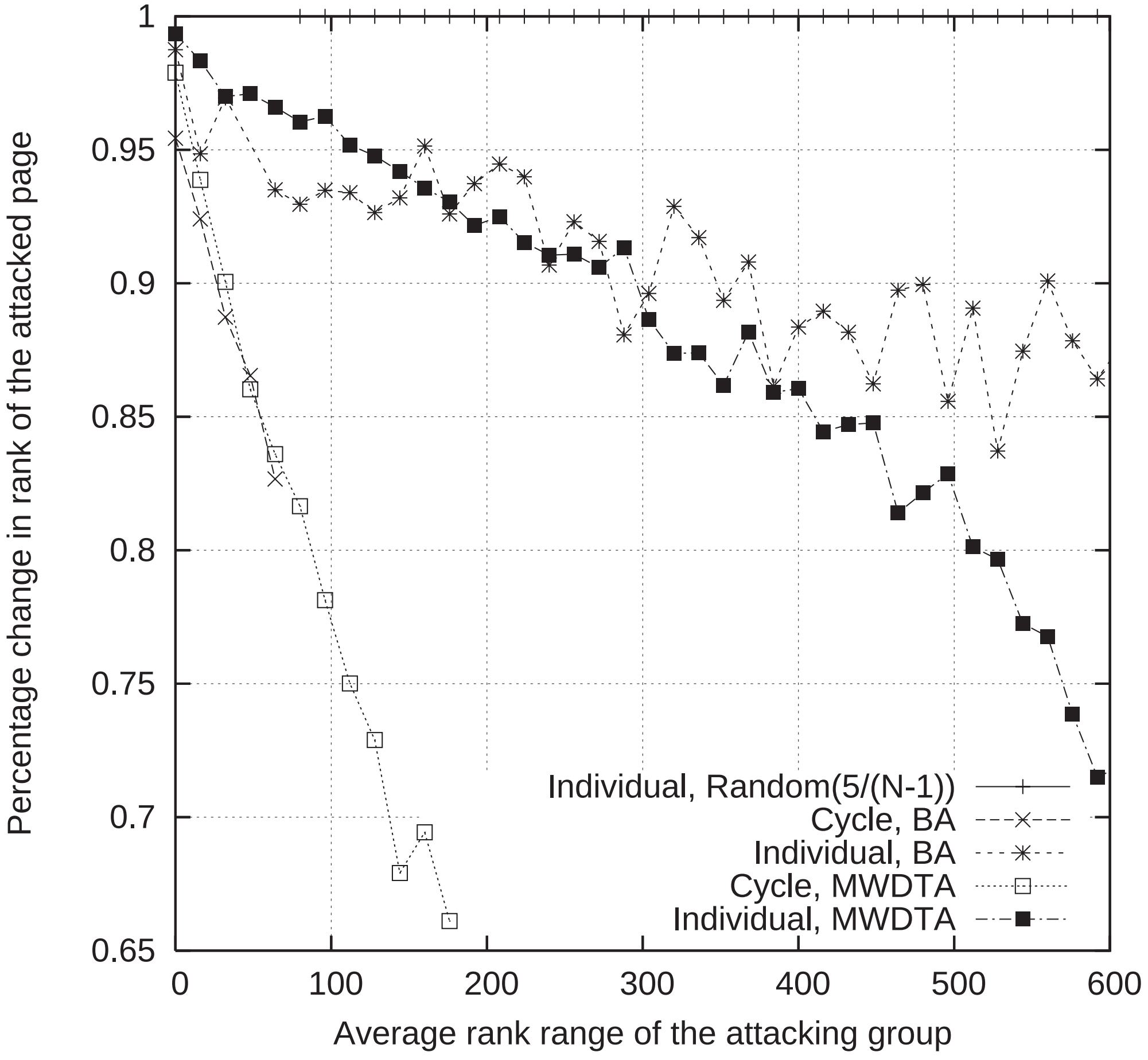

Figure 3: Experimental results on rank

and $\alpha$ is small. Our analysis has been focused on increasing a page's rank (pagerank manipulation) in the entire graph, i.e., the victims rank is increased for *every* query. The underlying model is that the query identifies a set of nodes (based on text and anchor text), which defines an induced subgraph of the original graph. However, the nodes are ranked according to pagerank in the original graph. This model has the feature that pageranks do not need to be recomputed for the specific query. An alternative approach is to order the nodes with respect to the pageranks in the induced subgraph (hence these pageranks would need to be recomputed for every query). Such a model would mean that one attempts to boost the pagerank with respect to a specific query and not others. Our analysis does not apply to this model, and it is no longer true that the optimal attack is the direct individual attack. The following example (with a single attacker) illustrates:

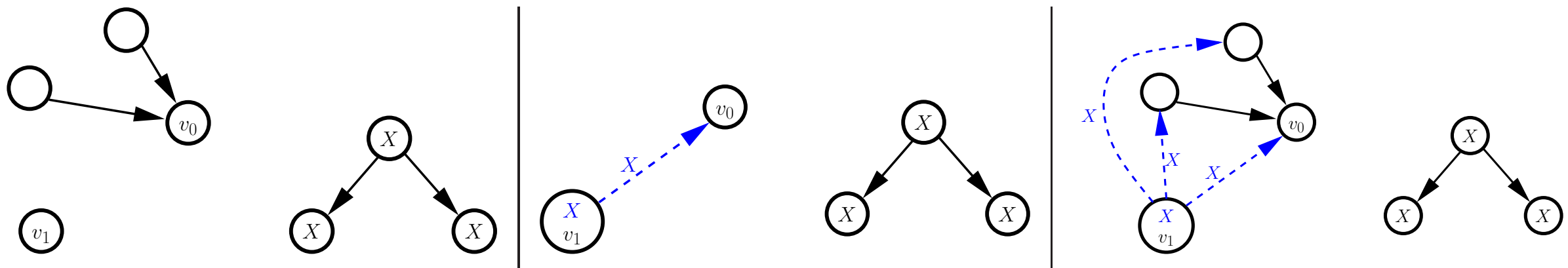

(a) Original graph. (b) Direct attack. (c) Indirect attack.

In (a) we show the original graph, where $X$ will be the query text and the attacker wants to boost the rank of $v_0$ with respect to $X$. In (b) we show the subgraph induced by the direct attack, where the attacker places $X$ in its page as well as in the anchor text of the link. In the resulting induced subgraph, the rank of $v_0$ is not the highest. The benefit of the non-direct attack in (c) is that other nodes that point to $v_0$ get included into the induced subgraph. Thus while the flow of rank from $v_1$ to $v_0$ is decreased, this is more than compensated for by the additional rank contribution from the newly included nodes. A better attack would arise if $v_1$ added another link to $v_0$. In fact for any attack in which $v_1$ has $k$ links to $v_0$, a strictly better attack with $k+1$ links is possible. In this example, there is no optimal attack. In general, we can formulate this notion by saying that the attacker should add the minimum number of links to all nodes with paths to the victim which do not contain the query text, and hence would not be included in the subgraph. The attackers should then place as many parallel direct links to $v_0$ as is feasible. The end effect is to include all nodes with paths to the victim with a minimum diversion of page rank. Of course, such a huge attack is not very practical, and an interesting question is to consider the optimal attack under this model when each attacker has a fixed budget of links.

The PageRank algorithm favors attacks from groups that are not well connected, which makes it harder to detect the attack, and accountability in such an attack formation becomes an issue: who is responsible for the attack? Different variations of the PageRank algorithm may suffer a similar fate if they propagate the pagerank in a similar way (for example Topic-Sensitive PageRank [15], provided that the attacking group is considered relevent to the query). In order to avoid such a fate (a dilemma faced by any ranking method open to manipulation by small groups), either one must change the ranking function or somehow exclude the attacking group from the search engine's database. While such an approach is a reasonable way to deal with private companies attempting to manipulate rankings based on their own views, it is not very democracy-friendly to arbitrarily remove certain pages from a search engine.

As discussed in [12], the PageRank algorithm makes certain assumptions about the user navigation patterns and the web structure that may not apply to the Web anymore. [12] considers the effect of dangling nodes in the pagerank computation and provides methods to adjust for them. They also point out that users will rarely (if ever) navigate to one of several billion pages uniformly – they may not even know that these pages exist. In fact, users generally start from known sites and navigate from there. Hence, random navigation is more likely to bring them to one of these "anchor" sites. The HostRank algorithm [12] uses this assumption to choose a set of anchor sites, and they show that such an approach is more resistant to attacks. Trustrank algorithm [14] uses a set of trusted pages to bias the random jump probability. An interesting problem would be to check whether the selection algorithm for trusted pages can be manipulated (if it is not fully manual). For example, pages can exhibit trustworthy behavior to gain trust and then sell this influence for spam links. It would be interesting to study the sensitivity of the algorithm to various types of attacks.

A related issue is that of navigation along links from a site. One is more likely to trust a link on a highly ranked page, and one is more likely to follow a link to a highly ranked page. For example, it might be *much* more probable to follow one of the links from a search engine or a news Web site than a regular web page. The probability to navigate from a page in the PageRank algorithm is independent of a page's rank, and the link one selects to navigate is random. A plausible alternative is that the probability to navigate from a page should be proportional to the page's pagerank, and the probability to use a particular outgoing link is proportional to the pagerank of the destination page. Such a navigation model would lead to an equation (analogous to (1)) of the form

$$p_i = \kappa\alpha p_i \sum_{(v_j,v_i)\in E} \frac{p_j^2}{\sum\limits_{(v_j,v_k)\in E} p_k} + \frac{1-\alpha}{N}.$$

More effort could be spent on how the transition probabilities generally affect the pageranks and their manipulability. [12] discusses such issues for nodes with unknown outgoing links and [21] uses the amount of traffic flow through the nodes to model the transition probabilities. It would be interesting to see what the optimal attack with such ranking algorithms is. In short, objective methods for the selection of the anchor sites or more plausible navigation models deserves closer examination. One must also bear in mind (see for example [12]) that the computational complexity of the algorithm is also an important practical consideration for any ranking algorithm.

Other factors, which we do not study here, might be significant to the success of an attack. [11] argues that anchor text pointing to a page gives information ragarding the subject matter of that page, and relationships between different pages. For example, Google may consider both the pagerank and the frequency of keywords in links pointing to a page when computing the score of the page. Google bombs in the past used the same keywords when pointing to the attacked page, i.e., the bombing links were correlated in that they all had the same keywords, whereas in general, links pointing to a website would not display such a correlation. If some linear combination of these two factors is then used in the final score, it will favor attacks over the natural Web behavior. If some small group of sites use a specific keyword to point to a victim, it is unlikely that this groups's sites are unrelated, and one could (for example) add pseudo-links among these sites, since the expection would be that they participate in some group structure. As our results show, these pseudo-links will reduce the magnitude of the attack. One could go so far as to say that if after the addition of such pseudo-links in the graph, the pagerank distribution does not change significantly, then the ranking algorithm should be more resistant to manipulation.

The analysis of the optimal attack structure provides a new tool for looking at resistance to link manipulation. Such metrics and an understanding of optimal attack formations for other algorithms should be fruitful directions for future work.

## Acknowledgements

We are grateful to Mark Goldberg for his initial feedback on this work.

This research is continuing through participation in the Network Science Collaborative Technology Alliance sponsored by the U.S. Army Research Laboratory under Agreement Number W911NF-09-2-0053. The views and conclusions contained in this document are those of the author(s) and should not be interpreted as representing the official policies, either expressed or implied, of the U.S. Army Research Laboratory or the U.S. Government. The U.S. Government is authorized to reproduce and distribute reprints for Government purposes notwithstanding any copyright notation hereon.

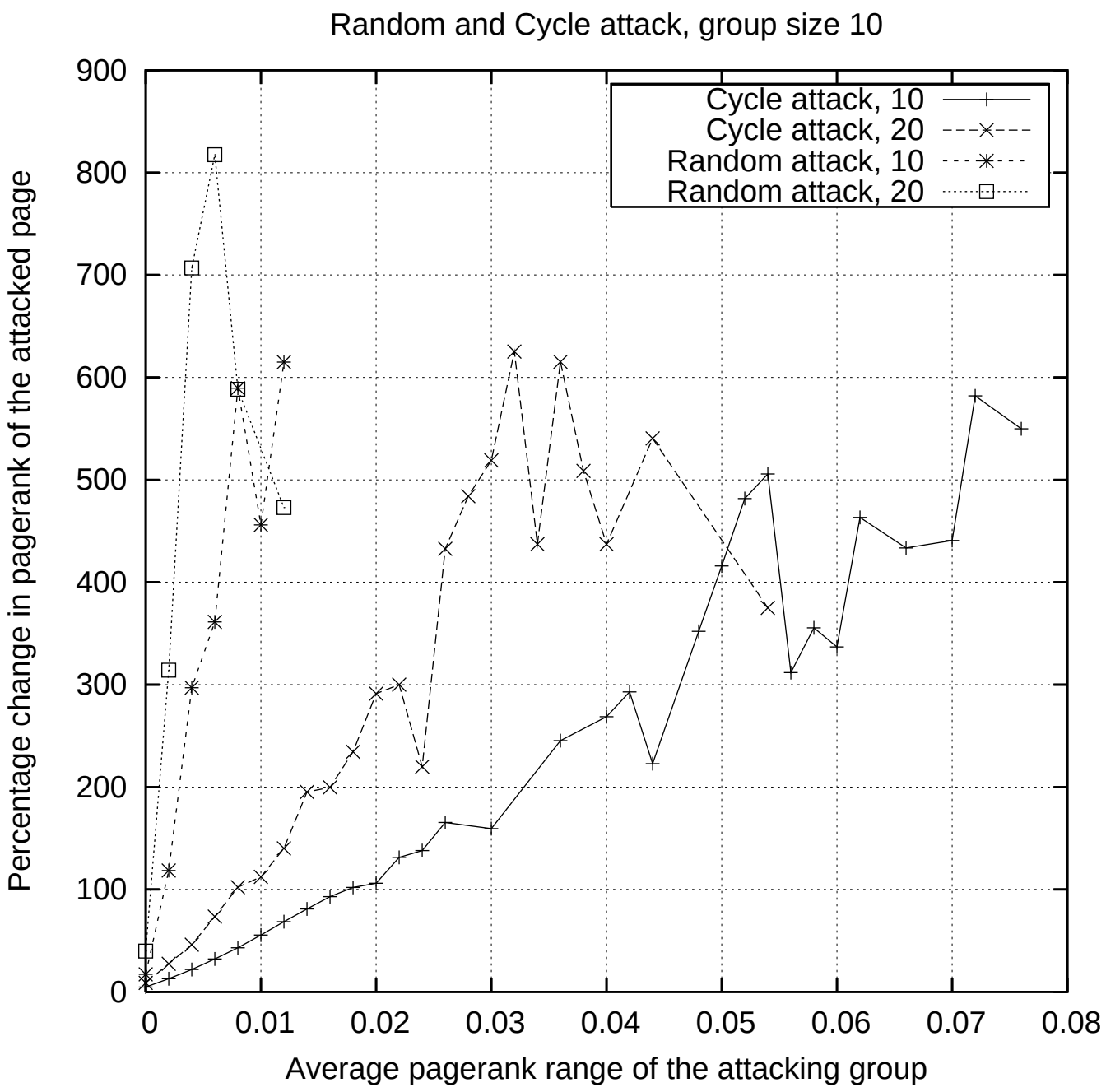
Random and Cycle attack, group size 10
Percentage change in pagerank of the attacked page
Average pagerank range of the attacking group
Cycle attack, 10
Cycle attack, 20
Random attack, 10
Random attack, 20
0
100
200
300
400
500
600
700
800
900
0.01
0.02
0.03
0.04
0.05
0.06
0.07
0.08

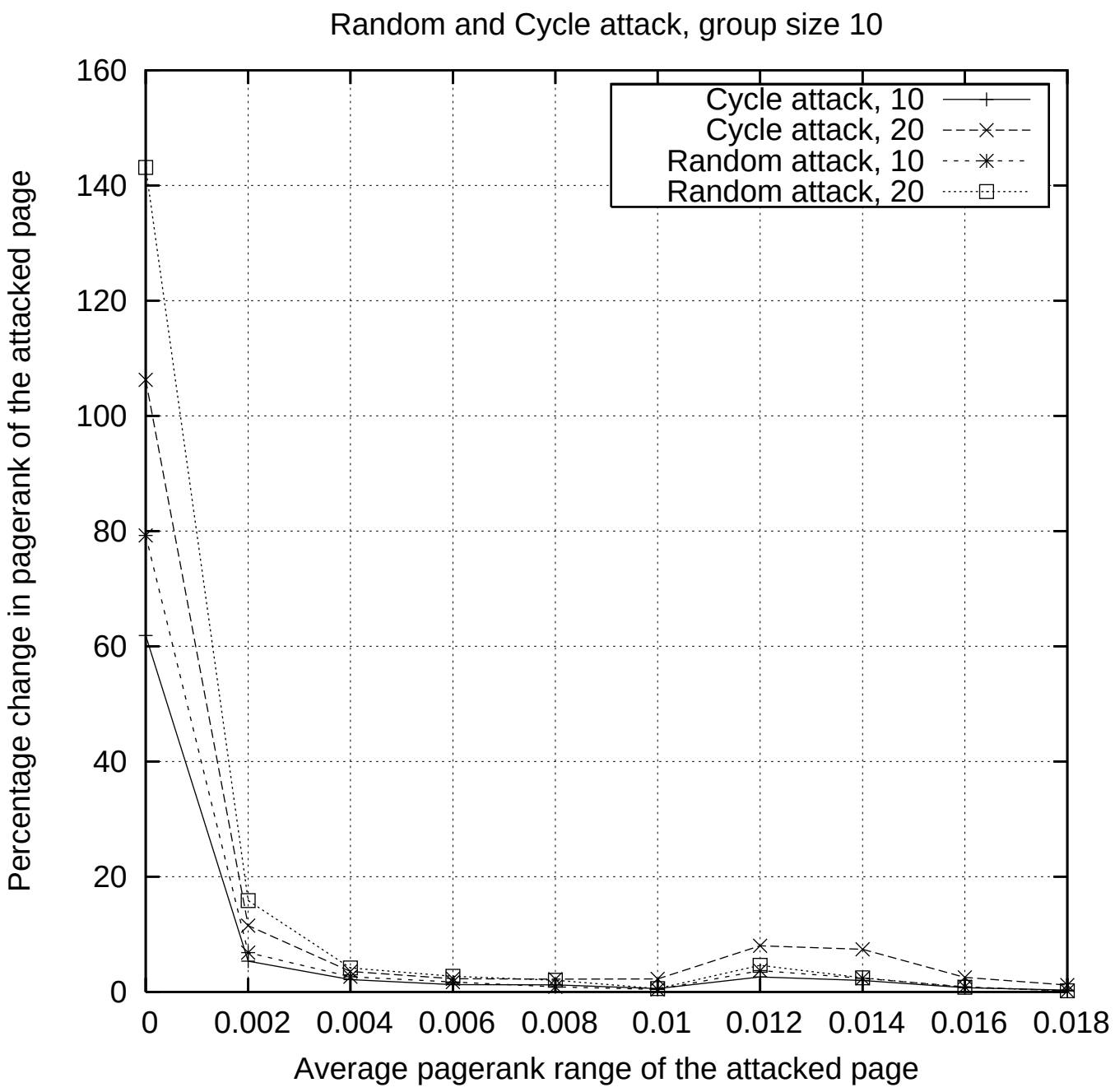
Random and Cycle attack, group size 10
Cycle attack, 10
Cycle attack, 20
Random attack, 10
Random attack, 20
Percentage change in pagerank of the attacked page
Average pagerank range of the attacked page
0
20
40
60
80
100
120
140
160
0.002
0.004
0.006
0.008
0.01
0.012
0.014
0.016
0.018

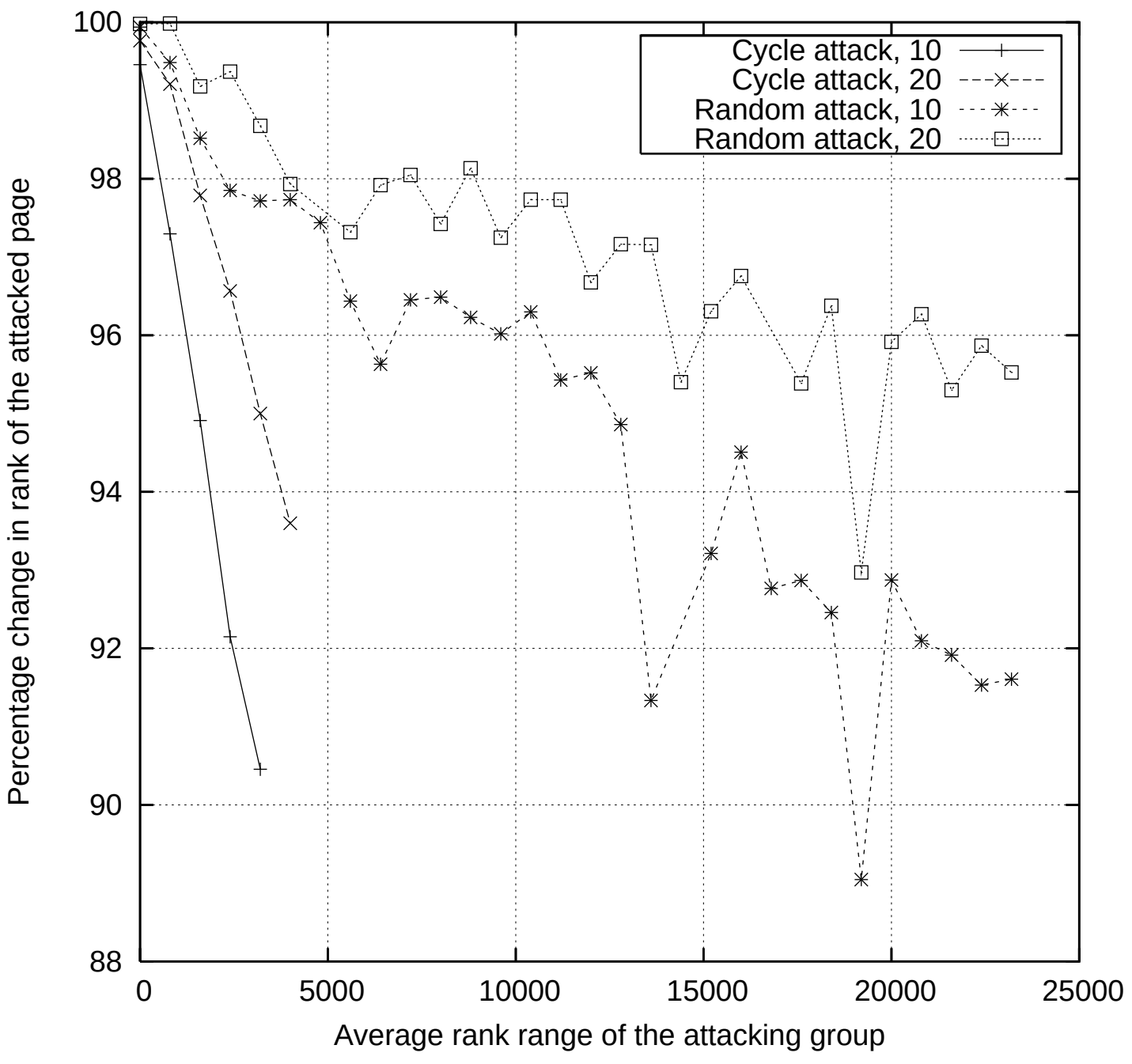

Cycle attack, 10
Cycle attack, 20
Random attack, 10
Random attack, 20
Percentage change in rank of the attacked page
Average rank range of the attacking group
100
98
96
94
92
90
88
0
5000
10000
15000
20000
25000

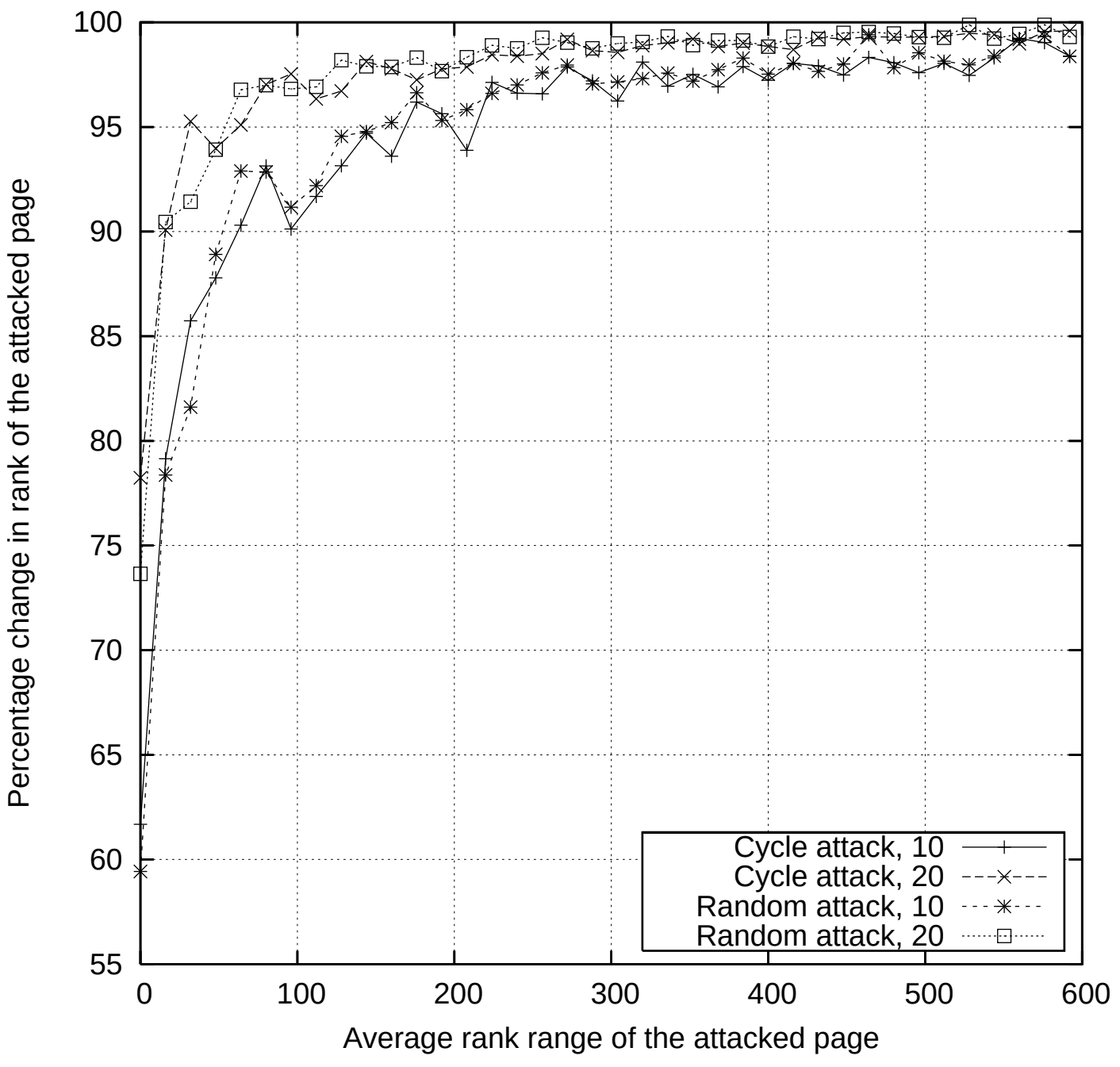

Percentage change in rank of the attacked page
Average rank range of the attacked page
Cycle attack, 10
Cycle attack, 20
Random attack, 10
Random attack, 20
100
95
90
85
80
75
70
65
60
55
0
100
200
300
400
500
600